\documentclass[a4paper,11pt]{article}
\pdfoutput=1 
\usepackage{jheppub}
\usepackage[T1]{fontenc} 
\usepackage[utf8]{inputenc}
\usepackage{amsmath}
\usepackage{xcolor}
\usepackage{amssymb}
\usepackage{mathtools}
\usepackage{amsfonts}
\usepackage{bbold}
\usepackage{enumerate}
\usepackage{tikz}
\usetikzlibrary{angles,quotes}
\usepackage{graphicx}
\usepackage{caption}
\usepackage{subcaption}
\usepackage{float}
\usepackage{ulem}
\usepackage{hyperref}
\usepackage{color}
\usepackage{tikz-network}
\usepackage{tensor}
\usetikzlibrary{arrows}

\begin{document}
\title{Spontaneously interacting qubits from Gauss-Bonnet}
\author{Sean Prudhoe,}

\emailAdd{stp5134@psu.edu}
\author{Rishabh Kumar,}
\emailAdd{rzk294@psu.edu}
\author{Sarah Shandera}
\emailAdd{ses47@psu.edu }
\date{April 2022}
\affiliation{Institute for Gravitation and the Cosmos, The Pennsylvania State University, 321 Whitmore, University Park, USA}
\affiliation{Department of Physics, The Pennsylvania State University, Davey 104 ,University Park, USA}

\abstract{
 Building on previous constructions examining how a collection of small, locally interacting quantum systems might emerge via spontaneous symmetry breaking from a single-particle system of high dimension, we consider a larger family of geometric loss functionals and explicitly construct several classes of critical metrics which "know about qubits" (KAQ). The loss functional consists of the Ricci scalar with the addition of the Gauss-Bonnet term, which introduces an order parameter that allows for spontaneous symmetry breaking. The appeal of this method is two-fold: (i) the Ricci scalar has already been shown to have KAQ critical metrics and (ii) exact equations of motions are known for loss functionals with generic curvature terms up to two derivatives. We show that KAQ critical metrics, which are solutions to the equations of motion in the space of left-invariant metrics with fixed determinant, exist for loss functionals that include the Gauss-Bonnet term. We find that exploiting the subalgebra structure leads us to natural classes of KAQ metrics which contain the familiar distributions (GUE, GOE, GSE) for random Hamiltonians. We introduce tools for this analysis that will allow for straightfoward, although numerically intensive, extension to other loss functionals and higher-dimension systems.}

\maketitle
  
\section{Introduction} 
 

Although the notion of emergent gravity has been studied from a variety of perspectives\footnote{A non-exhaustive list of examples includes \cite{Ishibashi:1997,Konopka:2008, Hu:2009jd,Hamma:2010,Verlinde:2010hp,Swingle:2014}.}, a simpler but still illuminating question to consider is the emergence of locality. A formal structure to pose that question, at least for toy systems described by large, finite-dimensional quantum systems, was recently suggested by Freedman and Zini \cite{Freedman:2021a,Freedman:2021b,Freedman:2021c}. They introduced functionals of geometry on the evolutionary operators of high-dimensional quantum systems and asked whether the geometries that minimize those functionals correspond to dynamics of a many-body quantum system with a notion of local interactions. The tools to answer this question are inner products on the Lie algebra $\mathfrak{su}(N)$, corresponding to left-invariant metrics on the associated manifold of the Lie group $SU(N)$ \cite{kobayashi1996foundations, Caselle2004,Gallier:2020}. Those inner products, or metrics, can be used to construct probability distributions over Hamiltonians that give preference to different classes of dynamics on a Hilbert space of dimension $N$. 



Here, we expand in two technical ways on the examples considered in the original statement of this program \cite{Freedman:2021a}. First, we consider a new set of geometrically motivated loss functionals which have critical points corresponding to a qubit structure decomposition. Reference \cite{Freedman:2021a} labeled critical points of that type as KAQ since they `know about qubits'. Motivated by the critical point structure of the Ricci scalar, we consider loss functions built from higher-order curvature terms. The exact equations of motion for such actions already exist in the literature \cite{Berger:1970,Muto:1974,Muto:1977, Patterson}. Secondly, we provide a construction for classes of KAQ metrics that generalize those recovered in \cite{Freedman:2021a} and originally found in \cite{Jensen:1971}. We use this construction as an ansatz for critical points of our loss functionals. In this way we are able to determine potential KAQ critical points in the space of our ansatz metrics, which then may be checked against the equations of motion. We do not need to search in the full space of left-invariant metrics, we only need to search in the exponentially reduced space of our ansatz KAQ metrics. Although still numerically intensive, this is a promising approach to apply to systems larger than those we treat in this article. 


In the rest of this section, we lay out in more detail the statement of the problem and the tools to be used, including the new loss functionals. Section \ref{sec:KAQparam} then describes in detail the parameterizations of KAQ metrics we find useful. In Section \ref{sec:EquationsOfMotion} we present the equations of motion that must be solved to find critical points and apply the parameterizations to find new critical points, presented in Section \ref{sec:results}. We conclude with implications and further directions in Section \ref{sec:Conclude}.


\subsection{Distributions over Hamiltonians} 
Rather than searching for some dynamics that picks out a specific Hamiltonian, it is natural to ask for a distribution over Hamiltonians that assigns a higher likelihood to a particular class with interesting behavior. A simple and familiar choice for a distribution over dynamics of a quantum system of dimension $N$ is the Gaussian Unitary Ensemble \cite{wigner1955,dyson1962statistical}, where each independent real number of the $N\times N$ Hermitian matrix that defines the Hamiltonian, $H$, is independently drawn from a Gaussian distribution. Equivalently,
 \begin{equation}
    \rho_{\rm GUE}\big(H\big)\propto e^{-{\rm tr}\big(H^{2}\big)}\,.
\end{equation}
This distribution is invariant under unitary transformations, $H\rightarrow UHU^{\dagger}$ (as is clear from the cyclic property of the trace). It does not give preference to Hamiltonians that have the many-body local structure typical of spin systems frequently observed in nature, where the many entries in the large matrix $H$ that correspond to interactions among nearly all the degrees of freedom would be suppressed. A different distribution that would support a many-body local structure when $N=2^d$ would, for example, have a basis constructed from Pauli words for qubits, $\otimes_{i=1}^d\sigma_i^{(J)}$, where each $\sigma^{(J)}$ is any of the Pauli matrices or the $2\times2$ identity. A choice of weights assigned to operators in this basis could define a distribution favoring some subset of operators, for example words of shorter length (measured by the number of non-identity Pauli matrices). 

To formalize distributions that are different from GUE, consider the set of inner products on the space of operators, which for quantum systems with Hilbert space of dimension $N$ is the algebra $\mathfrak{g}=\mathfrak{su}(N)$. The Lie algebra comes equipped with the Lie bracket, but the bracket itself does not assign an inner product. The GUE arises from the most symmetric choice of metric, given by the Killing form $K: \mathfrak{g}\times \mathfrak{g} \rightarrow\mathbb{C}$, which defines a map that is invariant under a basis change. Using $K$ defines the Killing-Cartan metric for $X$, $Y$ elements of the algebra
\begin{equation}
\delta(X,Y)\equiv {\rm tr}\big({\rm ad}X*{\rm ad}Y\big) \, ,
\end{equation}
where ${\rm ad}X$ is the map defined by ${\rm ad}X(Y)=[X,Y]$ for all $Y$. General inner products on the algebra correspond to left-invariant metrics on the group manifold for $G=SU(N)$, while the Killing-Cartan metric is special in that it is both left and right invariant i.e. bi-invariant. Although from the manifold point of view one can imagine more complex metrics that depend on some choice of coordinates, the set of left-invariant metrics is sufficient for constructing a larger class of distributions over Hamiltonians. The more general distributions over Hamiltonians can be written
\begin{equation}
    \rho_{\rm FZ}(\mathcal{O})\propto e^{-g(\mathcal{O},\mathcal{O})}
\end{equation}
where $g$ is a strictly left-invariant metric over $SU(N)$. To construct a left-invariant metric, one need only supply its value over the Lie algebra. This is equivalent to supplying the Lie algebra with an inner product i.e. 
\begin{equation}
    \langle\cdot , \cdot\rangle: \mathfrak{g}\times \mathfrak{g}\rightarrow \mathbb{C} 
\end{equation}
which may be propagated to the rest of the Lie group through use of the differential of left-multiplication \cite{kobayashi1996foundations}. Defining the metric in this way guarantees that it is indeed left-invariant. For a left-invariant metric to be bi-invariant, its associated inner product must be ad$\mathfrak{g}$ invariant i.e. 
\begin{equation}
\label{eq:bi-invariant}
    \langle [\mathcal{O},A],[\mathcal{O},B]\rangle =  \langle A,B\rangle
\end{equation} holds for all $\mathcal{O},A,B \in \mathfrak{su}(N) $. In Section \ref{sec:KAQparam} we construct several classes of strictly left-invariant metrics, and one may demonstrate they are strictly left-invariant by finding any particular $\mathcal{O}$ for which Eq.\ref{eq:bi-invariant} fails to hold.

In fact the Killing-Cartan geometry is the only bi-invariant geometry over SU(N) (up to overall scale) and so considering strictly left-invariant geometries allows for more structure. Left-invariant metrics can be distinguished using their principal axes, which correspond to the elements of the orthonormal basis $(X_{a})$ of an associated inner product. These operators are orthonormal in the sense that $\langle X_{a},X_{b}\rangle=\delta_{ab}$ where $\delta_{ab}$ is the Kronecker delta and the indices take values between 1 and $4^{d}-1$. In general, the orthonormal basis of an inner product produces a non-holonomic or non-coordinate basis over the manifold since the structure constants ($C\indices{^c_{ab}}$ ) are non-zero. Recall that the structure constants of a basis are given by 
\begin{equation}
[X_{a},X_{b}]=\sum_{c}C\indices{^c_{ab}}X_{c}\,.
\end{equation}
Therefore, one may not use the standard formulas from general relativity when computing curvature functionals. For example, in a non-holonomic basis the Christoffel connection picks up a term proportional to the structure constants. See Appendix \ref{app:functionals} for all relevant formula of the curvature functionals expressed in a non-holonomic basis. 
 
\subsection{Many-body dynamics via preferred geometries} 
Although one can construct any distribution over Hamiltonians by hand, it is interesting to ask if there may be a geometrical means of picking out interesting classes, which perhaps could be dynamically realized as a spontaneous symmetry breaking process that fragments a large "single-particle" quantum system into an ensemble of small quantum systems interacting in a way that resembles local, many-body physics. Freedman and Zini \cite{Freedman:2021a,Freedman:2021b,Freedman:2021c} considered a family of functionals of the group geometry, parameterized by the metric and the structure constants, and explored whether the minima of these functionals select out KAQ metrics. 

Here, we explore this idea a bit further by looking at functionals $\mathcal{L}[g]$ that are natural on the group manifold, defined by the set of two-derivative curvature tensors 
 \begin{equation}
 \label{eq:LossFunctionals}
 \begin{split}
     \mathcal{L}[g]&=R+\alpha R^{2}+\beta R_{ab}R^{ab}+\gamma R_{abcd}R^{abcd} \\
     &=R+\alpha\mathcal{R}_{0}+\beta \mathcal{R}_{2}+\gamma \mathcal{R}_{4}\, .
\end{split}
\end{equation}
The critical points of just the Ricci scalar, $R$, were first studied by \cite{Jensen:1971}, and the same results were recovered by \cite{Freedman:2021a}. The larger class of functionals in Eq.(\ref{eq:LossFunctionals}) is especially tractable to study since the conditions for a metric to be a critical point are already known \cite{Patterson,1977493}. We return to this in Section \ref{sec:EquationsOfMotion}.
While \cite{Freedman:2021a} used numerical techniques to find all critical points of some loss functionals and then determine if they corresponded to KAQ or non-KAQ metrics, we take a different approach and instead explore whether or not KAQ metrics occur as critical points (and ideally minima) of an expanded set of loss functions. To do so, we next introduce parameterizations of KAQ metrics, including generalizations of those that correspond to the KAQ critical points found in \cite{Jensen:1971} for $\mathcal{L}[g]=R$. This allows us to explore the structure and properties of the KAQ metrics in more detail, although with the drawback that we cannot determine the relative frequency of KAQ vs non-KAQ minima.

\section{KAQ parameterization schemes}
\label{sec:KAQparam}
Among the $N+1\choose 2$ distinct metrics on $SU(N)$, only some will have a structure that is compatible with a tensor product decomposition into $d$ qubit operators when $N=2^d$. This notion can be expanded to apply to tensor decompositions of more general large $N$ spaces that can include factors of dimension other than two (qudits) \cite{Freedman:2021b,Freedman:2021c}. The KAQ property of a metric $g$ is decoded from its (possibly non-unique) principle axes. An observable $E_{a}$ is said to be a principle axis of the metric if
\begin{equation}
\label{eq:PA}
g\indices{^a_b}E_{a}=\lambda_{b}E_{b}
\end{equation}
that is $E_{a} $ is an eigenvector of the tensor $g\indices{^a_{b}}=g_{cb}\delta^{ca}$, where $\delta_{ab}$ is the Killing-Cartan metric. The location of indices in the previous equation matter; they are chosen so that we may compute eigenvectors. Later on we compute only the components of tensors in an orthonormal basis of the metric, therefore placement of indices becomes essentially irrelevant. In Eq.\ref{eq:PA} the index placement is relevant, as we require a tensor that maps vectors to vectors.

The principal axis $\{E_{a}\}$ need only be orthogonal, hence the need for a distinct from an orthonormal basis which are denoted by $\{X_{a}\}$. A left-invariant (but not bi-invariant) metric is said to be KAQ if $\mathbf{a}$ basis of principal axis exist such that
\begin{equation}
\begin{split}
&\Phi[E_{a}]=\sigma^{(a_{1})}_1\otimes...\otimes \sigma_d^{(a_{d})} \\
&[E_{a},E_{b}]=\mathcal{P}\indices{^c_{ab}}E_{c}
\end{split}
\end{equation}
where $\Phi$ is a Lie algebra isomorphism i.e. a bijective linear map which preserves commutation relations and the tensors $\mathcal{P}\indices{^c_{ab}}$ are the structure constants of $\mathfrak{s}\big(\mathfrak{u}(2)^{\otimes d}\big)$, the Lie algebra $\mathfrak{u}(2)^{\otimes d}$ with the generator $\mathbb{1}^{\otimes d}$ removed. While our definition is slightly different than that given in \cite{Freedman:2021a}, they are in fact equivalent. Taking a given Pauli word to a linear combination requires using unitary conjugation which will not effect commutation relations. It is important to stress that the metric will typically have many possible bases of principle axes, and only one needs to satisfy the KAQ condition. Furthermore, among that restricted set of KAQ metrics, not all will generate many-body local dynamics by suppressing the contributions from Pauli words with length close to $d$. 

The degeneracy pattern of metric determines the freedom there is in choosing a KAQ basis. In the case of no degenerate eigenvalues, the metric is KAQ iff all principal axes already align with some Pauli word basis. This is clearly a special case. More generally, a degenerate eigenspace may be decomposed into a basis that aligns with Pauli words, although some ``decoding" may be required. Decoding here means that the degenerate principal axes are mixed using an element of $SO(4^{d}-1)$ which is not an inner automorphism of $\mathfrak{su}(2^{d})$. Such transformations keep distinct degenerate axis orthogonal, but they do not preserve commutation relations. Later in this section we give an explicit example of such a decoding process, where we demonstrate that Gell-Mann matrices may serve the role of a KAQ basis for certain metrics.

An obvious structure to make use of in searching for KAQ metrics is the $\mathfrak{so}(2^{d})$ subalgebra of $\mathfrak{su}(2^{d})$, which one might expect contains a set of axes decodable to the length-one Pauli words, the single-qubit operators \cite{Khaneja:2000}. In the case of $\mathfrak{su}(4)$, the construction is simple: if one starts with the natural basis of Gell-mann matrices, the single-qubit operators can be recovered by identifying linear combinations of matrices in the sub-algebra that have a tensor product form and then making a rotation to align those linear combinations with length-one Pauli words.  On the other hand, the other obvious subalgebra decomposition, $\mathfrak{sp}(2^{d-1})$, contains degenerate subspaces that must align with Pauli word basis, and others which may require decoding. Operationally, we consider this case in detail below to illuminate the relationship between known critical points of the Ricci scalar, which carry this subgroup structure, and critical points with KAQ structure.


In order to construct parameterizations of KAQ metrics, which can then be used as ansatze for critical metrics, we use the fact that a left-invariant metric over a Lie group is entirely specified by an inner product over the corresponding Lie algebra. In this way instead of referring to the metric, one may just as well refer to an orthonormal basis of the associated inner product. For excellent reviews of the Lie algebra and geometric background needed for these constructions, see \cite{kobayashi1996foundations, Caselle2004,Gallier:2020}.  

\subsection{Riemannian geometry of compact symmetric Lie algebras}
Consider first the Killing-Cartan geometry naturally available to SU($4$). It is the unique (up to scale) bi-invariant geometry over the special unitary group, and describes color dynamics (in 0+1 dimensions) mediated by 15 gluons. The generalized Gell-Mann matrices, $G_a$ provide an orthonormal basis, where we use the definitions provided in \cite{Bossion:2021zjn}. However, we include an additional factor of $i$ in our construction of generalized Gell-Mann matrices therefore taking them to be anti-Hermitian operators. This is more in keeping with standard notation in differential geometry. We further consider a non-standard normalization 
\begin{equation}
    {\rm tr}\big(G_{a}^{\dag}G_{b}\big)=\frac{1}{2}\delta_{ab}\,,
\end{equation}
which will enter the structure constants
\begin{equation}\label{eq: GellMannCommutation}
[G_{a},G_{b}]=K\indices{^c_{ab}}G_{c} \, .
\end{equation}
Denoting $K_{a}$ as the matrix of structure constants with entries $K\indices{^c_{ab}}$ in the Gell-Mann basis we then have 
\begin{equation}
{\rm tr}K_{a}K_{b}=-4\delta_{ab}
\end{equation}
and bi-invariance implies that $K^{\rm T}_{a}=-K_{a}$ i.e. the structure constants are totally anti-symmetric. This equation is negative definite due to the compactness of the special unitary group. The normalization choice here leads to a convenient normalization later on in the loss functionals, where the Ricci scalar takes the value $R=15=N^{2}-1$ when evaluated over the Killing-Cartan geometry.

The Killing-Cartan geometry serves the role of a simple fiducial metric, as well as the assumed unstable starting point for spontaneous symmetry breaking. Instead of studying a more complicated left-invariant metric directly, one may look at the linear transformation ($\omega$) relating one of its orthonormal bases to that of the Killing-Cartan geometry \cite{Jensen:1971}. The transformation $\omega$ is required to fix the determinant, which we impose in order to use the equations of motion referenced in Section \ref{sec:EquationsOfMotion}. The linear transformation is related to the metric when expressed in the Gell-Mann basis as $g=\omega^{-2}$. One may check such transformation certainly maps Gell-Mann letters to an orthonormal basis of $g$.

Assume we have related a new orthonormal basis to the Gell-Mann matrices by a special linear transformation $\tilde{G}_{a}=\sum_{a}\omega_{ab} G_{b}$. Then the structure constants are related by
\begin{equation}
\begin{split}
&\tilde{K}_{a}=\sum_{b}\omega_{ab}[\omega^{-1}K_{b}\omega]\,, \\ 
\end{split}
\end{equation}
where $K_b$ is the matrix with $kj$th entry $K^{k}_{bj}$. For these calculations the location of an index (up or down) is unimportant, as the metric is always an identity matrix. The preceding equation makes explicit the matrix multiplication that must be performed to determine the new structure constants. That is, the transformation law may be written equivalently as
\begin{equation}
\tilde{K}\indices{^c_{ad}}=\tilde{K}_{cad}=\sum_{bef}\omega_{ab}\omega^{-1}_{ce}K_{ebf}\omega_{fd} \,.
\end{equation}
The utility is that now curvature functions may be expressed in terms of Killing-Cartan tensor networks, where the total anti-symmetry can be leveraged. 

We shall end with a final bit of notation, which simplifies the construction of KAQ metrics. As the metrics are considered over the Lie algebra, it is helpful to use the adjoint representation (${\rm ad}\mathfrak{g}$). Here observables themselves become operators acting over $\mathfrak{g}$. For example, we can define kets from the Gell-mann matrices $G_{a}\in\mathfrak{su}(2^{d})$, denoted $|G_{a}\rangle$. We then  define dual vectors with respect to the Killing-Cartan metric i.e. 
\begin{equation}
\langle G_{b}|\equiv 2{\rm tr}\left[G_{b}^{\dag}(\cdot)\right]\,.
\end{equation}
 Defined in this way $\{|G_a\rangle\}$ is an orthonormal basis of $\mathfrak{g}$, with inner product $\langle G_a|G_b\rangle = \delta_{ab}$. Projectors are defined in the standard way
\begin{equation}
\Pi_{a}=|G_a\rangle\langle G_a| \,,
\end{equation}
and the action of the observables becomes
(\ref{eq: GellMannCommutation})
\begin{equation}
{\rm ad}G_{a}|G_{b}\rangle=K_{a}|G_{b}\rangle = |[G_{a},G_{b}]\rangle = \sum_{c}K_{cab}|G_{c}\rangle\,.
\end{equation}
We also obtain a matrix representation for the action of any metric over $\mathfrak{g}$. For example the linear transformation $\omega$ takes the form
\begin{equation}
\omega=\sum_{ab}\omega_{ab}|G_{a}\rangle\langle G_{b}|
\end{equation}
where $\omega_{ab}=\omega_{ba}$ and $\omega_{ab}\in \mathbb{R}$.

\subsection{Jensen geometries over SU(4)}
In \cite{Jensen:1971} the author searched for the critical points of the Ricci scalar curvature ($R$) in the space of left-invariant metrics with fixed determinant. It was shown that for unimodular Lie algebras (${\rm tr}\left[C_{a}\right]=0$), Einstein metrics are precisely the critical points of $R$. While the proof was not constructive for general unimodular Lie groups, by specializing the author was able to construct several classes of Einstein metrics, now called Jensen metrics. These were found over symmetric Lie algebras i.e. those Lie algebras with at least one Cartan decomposition. The Cartan decomposition plays a crucial role in forming the ansatz metric used in \cite{Jensen:1971} to find Einstein metrics. Further exploring the algebraic structures of these decompositions is an interesting point we return to later. We find it useful to first explore the algebraic properties of Jensen metric through explicit constructions of two classes relevant for this work.

Further assuming that the manifold is compact guarantees the existence of a strictly left-invariant Einstein metric. The Lie algebra $\mathfrak{su}(4)$ is compact and symmetric. Thus it may be Cartan decomposed into a subalgebra and its orthogonal complement with respect to the Killing-Cartan form, allowing for non-trivial Ricci critical metrics. There are two inequivalent ways to do this, corresponding to the two non-isomorphic subalgebras $\mathfrak{so}(4)$, and $\mathfrak{sp}(2)$ the compact symplectic Lie algebra. 

The heart of the Cartan decomposition is to break the Lie algebra into the subalgebra and the orthogonal complement with respect to the Killing-Cartan form. That is, given a subalgebra $\mathfrak{R}\subset \mathfrak{g}$ we decompose $\mathfrak{g}$ as
 \begin{equation}
\mathfrak{g}=\mathfrak{R}\oplus\mathfrak{M} 
 \end{equation}
 where $\delta(A,B)=0$ for all $A\in\mathfrak{R}$ and $B\in\mathfrak{M}$. We say the pair $(\mathfrak{g},\mathfrak{R})$ forms a Cartan decomposition if the following are satisfied
\begin{equation}
    \begin{split}
[\mathfrak{R},\mathfrak{R}]\subseteq \mathfrak{R}\,, \quad  [\mathfrak{R},\mathfrak{M}]\subseteq \mathfrak{M}\;\;, {\rm and} \quad
[\mathfrak{M},\mathfrak{M}]\subseteq \mathfrak{R} \, .
    \end{split}
    \end{equation}
The commutation relations above are equivalent to the existence of a certain kind of Lie algebra isomorphism, known as the Cartan involution ($\theta$). Explicitly  a Cartan involution is a Lie algebra automorphism $\theta:\mathfrak{g}\rightarrow \mathfrak{g}$ such that \cite{helgason2001differential}
\begin{equation}
    \begin{split} 
\theta=\begin{bmatrix}
    \mathbb{1}_{r} & 0\\ 0&-\mathbb{1}_{m} 
\end{bmatrix}
    \end{split}
\end{equation}
where $r={\rm dim}\mathfrak{R}$ and $m={\rm dim}\mathfrak{M}$. In what follows we construct the two types of Jensen metrics that exist for $\mathfrak{su}(4)$. To do so, we construct the transformation $\omega$, which takes an orthonormal basis of the Killing-Cartan metric to an orthonormal basis of the given critical metric. These transformations are generated by the symmetric, traceless operator
\begin{equation}
\label{eq:BJensen}
    B= \begin{bmatrix}
        \frac{1}{r}\mathbb{1}_{r} & \mathbf{0} \\ \mathbf{0} & -\frac{1}{m}\mathbb{1}_{m}
    \end{bmatrix}
\end{equation}
yielding the linear transformation
\begin{equation}
\label{eq:omegaJensen}
 \omega=\exp{\tau B} \,.
\end{equation}
The metric corresponding to $\omega(\tau)$ is a non-trivial critical point of the Ricci scalar iff
\begin{equation}
\label{eq:tauJensen}
    \tau= \frac{rm}{2(r+m)}\ln{\frac{2r+m}{2r-m}} \,.
\end{equation}
 These metrics are KAQ due to the following. First, the subalgebras which are pulled out for these two Cartan decompositions are generated by known sets of Pauli-words. This in conjunction with the large degeneracy in the Cartan blocks that ensures a KAQ basis may be defined. We do this explicitly in the next section. 
 
 Before moving to our more general sets of KAQ parameterizations, it is worth explaining the algebraic structures we leverage. To begin we simply note that the Jensen metrics have a reduced symmetry structure compared to the Killing-Cartan metric. Using the form of Jensen metrics, defined by Eq.(\ref{eq:BJensen})-Eq.(\ref{eq:tauJensen}), and the commutation relations, one may show that the Jensen metrics are only bi-invariant with respect to $\mathfrak{R}$. Thus since Jensen metrics are not ad$\mathfrak{g}$ invariant, they cannot support GUE dynamics. However the ad$\mathfrak{R}$ invariance allows for the creation of Gaussian ensembles over smaller sets of matrices. For example the $\mathfrak{so}(4)$ Jensen metric gives dynamics via the Ricci scalar where the GUE spontaneously breaks down to a model which approximates a lower dimensional Gaussian orthogonal ensemble (GOE),
\begin{equation}
   \rho= \rho_{\rm GUE}(H^{(N^2-1)}) \rightarrow \rho\approx\rho_{\rm GOE}(H^{(N(N-1)/2)})\,.
\end{equation}
It does not exactly yield the GOE as there is a non-negligible probability for observables outside of the $\mathfrak{so}(4)$ subalgebra to contribute to the Hamiltonian. Similarly, the $\mathfrak{sp}(2)$ Jensen metrics generate an approximate Gaussian Symplectic Ensemble (GSE). 
 
 But there are other known examples of Einstein metrics over SU(N), which have more varied structure than Jensen metrics. Most of these metrics are naturally reductive like the Jensen metrics. Naturally reductive metrics are more general than Jensen metrics, but they remain ad$\mathfrak{R}$ invariant. The more varied structure is obtained by decomposing the Lie algebra further, isolating certain algebraic structures of $\mathfrak{R}$. Given a Lie algebra with an orthogonal (but not necessarily Cartan) decomposition, one further decomposes it as  
 \begin{equation}
\mathfrak{g}=\mathfrak{R}\oplus\mathfrak{M}=\mathfrak{C}\oplus\mathfrak{I}_{1}\oplus...\oplus \mathfrak{I}_{q}\oplus\mathfrak{M}
 \end{equation}
 where $\mathfrak{C}$ is the center of $\mathfrak{R}$ ([$\mathfrak{C},\mathfrak{R}$]=0) and  $\mathfrak{I}_{i}$($i>0$) are simple ideals satisfying  
 \begin{equation}
[\mathfrak{R},\mathfrak{I}_{i}]\subset\mathfrak{I}_{i} \,.
 \end{equation}
 Using this decomposition, a general form of naturally reductive metrics is given in \cite{DAtri:1979}, 
 \begin{equation}
\langle \,|\,\rangle=\langle\,|\, \rangle_{\mathfrak{C}}+\lambda_{1}\mathbb{1}_{\mathfrak{I}_{1}}+...+\lambda_{q}\mathbb{1}_{\mathfrak{I}_{q}}+\mu\mathbb{1}_{\mathfrak{M}}
 \end{equation}
where $\langle\,|\,\rangle_{\mathfrak{C}}$ is a general inner product over $\mathfrak{C}$, and $\lambda_{i}$ and $\mu$ are non-negative. The authors used these parameterization to find examples of naturally reductive (non-Jensen) Einstein metrics, which are also critical points of $R$. The critical points of $R$ found in \cite{Freedman:2021a, Freedman:2021b, Freedman:2021c} were only of Jensen type. We find evidence later for why the non-Jensen type critical points were not found during the evolutionary search done by Freedman and Zini.

 It has recently been shown that non-naturally reductive Einstein metrics exist over $SU(N)$ \cite{Arvanitoyeorgos:2020}, but we do not yet know how large the overlap is with KAQ metrics. With that said, we shall also consider metric parameterizations which are non-naturally reductive. We motivate our construction by considering metrics with a reduced bi-invariance compared with the Jensen metrics. That is we consider metrics which are bi-invariant 
 with respect to a Cartan subalgebra of $\mathfrak{R}$. Explicitly we decompose the Lie algebra as
\begin{equation}
\mathfrak{g}=\mathfrak{R}\oplus\mathfrak{M}=\mathfrak{R}_{0}\oplus\mathfrak{R}_{1}\oplus...\oplus \mathfrak{R}_{q}\oplus\mathfrak{M}
 \end{equation}
where $\mathfrak{R}_{0}$ is a Cartan subalgebra of $\mathfrak{R}$, and $\mathfrak{R}_{i}$ are subspaces satisfying 
\begin{equation}
[\mathfrak{R}_{0},\mathfrak{R}_{i}]\subset \mathfrak{R}_{i} \,.
\end{equation}
Using this decomposition we define the following metrics 
 \begin{equation}
\langle \,|\,\rangle=\langle\,|\, \rangle_{\mathfrak{R}_{0}}+\lambda_{1}\mathbb{1}_{\mathfrak{R}_{1}}+...+\lambda_{q}\mathbb{1}_{\mathfrak{R}_{q}}+\mu\mathbb{1}_{\mathfrak{M}}
 \end{equation}
 where $\lambda_{i},\mu>0$. It is simple to show these metrics are indeed ad$\mathfrak{R}_{0}$ invariant using the preceding commutation relations. 
\subsection{Cipher classes }
In this section we introduce our KAQ parameterizations, which we break into cipher classes. These classes are distinguished by the existence of non-isomorphic KAQ bases. A given cipher class will have a basis of principal axes that is mixed between Gell-Mann letters and Pauli words. The smallest of these classes are the untranslated KAQ (UKAQ) metrics. These metrics have a KAQ basis consisting entirely of Gell-Mann letters. Such metrics are the most difficult to decode the KAQ property. The next biggest class are the partially translated KAQ metrics (PKAQ), followed by the fully translated KAQ metrics (FKAQ). The FKAQ metrics only have KAQ bases consisting entirely of Pauli words, thus they require no decoding when checking the KAQ property. We have the inclusion relation 
\begin{equation}
    \mathbf{U}{\rm KAQ }\subset \mathbf{P}{\rm KAQ} \subset \mathbf{F}{\rm KAQ}\,.
\end{equation}
For what follows we assume the basis principal axes maps to pure tensor Pauli words i.e.
\begin{equation}
\Phi\left[\tilde{E}_{a}\right]=\sigma_{a_{1}}\otimes...\otimes\sigma_{a_{d}}
\end{equation}
where $\tilde{E}_{a}$ represents the bases of principal axes post possible decoding process. It is useful to introduce some additional notation for the generalized Gell-Mann matrices, splitting them into three groups  
\begin{equation}
    \{ G_{a}\}=\{\{iA_{l}\},\{iS_{l}\},\{iD_{p}\}\}
\end{equation}
which are the anti-symmetric matrices, the off-diagonal symmetric matrices, and the diagonal matrices respectively. For a generic $N$ these indices take values in  
\begin{equation}
    \begin{split}
        &1\leq l \leq {N\choose 2} \\
        & 1\leq p\leq N-1
    \end{split}
\end{equation}
How we label the Gell-Mann matrices follows the notation from \cite{Bossion:2021zjn}, although we use a different overall normalization.
\subsubsection{Untranslated KAQ metrics}
We consider first the class of untranslated KAQ metrics (UKAQ). As a natural example, we use generalizations of the $\mathfrak{so}(4)$ Jensen metrics that are considerably less degenerate. To begin the construction we choose an orthonormal basis of the Killing-Cartan metric. We define generalized $\mathfrak{so}(4)$ Jensen metrics as those with an orthonormal basis compatible with the involution $\theta[X]=-X^{\rm T}$, i.e. an orthonormal basis constructed from eigenvectors of $\theta$. Notice that
\begin{equation}
    \begin{split}
        &\theta[A_{l}]=A_{l} \\
        &\theta[S_{l}]=-S_{l} \\
        &\theta[D_{p}]=-D_{p}
    \end{split}
\end{equation}
thus Gell-Mann letters are a good fiducial basis that can be used to construct generalized $\mathfrak{so}(4)$ Jensen metrics. 

Now notice that the eigen-spectra of Gell-Mann letters do not match those of the Pauli words. Therefore it is impossible to relate these bases through unitary conjugation i.e. inner automorphism. But in order to serve as a basis of principal axes, a Lie algebra homomorphism to Pauli words must exist. See also \cite{Khaneja:2000, Earp:2005}. Having ruled out the existence of an inner automorphism (unitary conjugation) that accomplishes the translation to Pauli words, the only remaining isomorphisms are the outermorphisms. But the only non-trivial outermorphism of $\mathfrak{su}(n)$ is complex conjugation. Complex conjugation combined with unitary conjugation can never match the eigen-spectra of Gell-Mann letters and Pauli words.  

Therefore we need to be able to decode the Gell-Mann matrices for the constructed metrics to be KAQ. It is straightforward to show that they may serve as $\mathbf{a}$ basis of principal axis, so long as we assume certain degeneracy patterns exist in the metric. To elucidate this concept, we construct the following dictionary that translates Gell-Mann letters to Pauli words 
\begin{equation}
    \begin{split}
    & (A_{1}+A_{6})=\frac{1}{2}\big(\mathbb{1}_{1}\otimes Y_{2} \big),\quad (A_{1}-A_{6})=\frac{1}{2}\big(Z_{1}\otimes Y_{2}\big) \\
     & (A_{2}+A_{5})=\frac{1}{2}\big(Y_{1}\otimes \mathbb{1}_{2}\big), \quad (A_{2}-A_{5})=\frac{1}{2}\Big(Y_{1}\otimes Z_{2}\Big) \\
    & (A_{3}+A_{4})=\frac{1}{2}\Big(Y_{1}\otimes X_{2} \Big),\quad (A_{3}-A_{4})=\frac{1}{2}\Big(X_{1}\otimes Y_{2}\Big)\,. \\
    \end{split}
\end{equation}
We may further construct Pauli words from $\mathfrak{M}$
\begin{equation}
    \begin{split}
&(S_{1}+S_{6})=\frac{1}{2}\Big(\mathbb{1}_{1}\otimes X_{2}\Big), \quad (S_{1}-S_{6})=\frac{1}{2}\Big(Z_{1}\otimes X_{2} \Big)\\
     & (S_{2}+S_{5})=\frac{1}{2}\Big(X_{1}\otimes \mathbb{1}_{2}\Big), \quad (S_{2}-S_{5})=\frac{1}{2}\Big(X_{1}\otimes Z_{2} \Big)\\
    & (S_{3}+S_{4})=\frac{1}{2}\Big(X_{1}\otimes X_{2}\Big), \quad (S_{4}-S_{3})=\frac{1}{2}\Big(Y_{1}\otimes Y_{2}\Big) \\
    &\left(D_{1}-\sqrt{\frac{1}{3}}D_{2}+\sqrt{\frac{2}{3}}D_{3}\right)= \frac{1}{2}\Big(\mathbb{1}_{1}\otimes Z_{2}\Big) \\
    &\left(D_{1}+\sqrt{\frac{1}{3}}D_{2}-\sqrt{\frac{2}{3}}D_{3}\right)= \frac{1}{2}\Big(Z_{1}\otimes Z_{2} \Big)  \\ 
    &\left(0D_{1}+\sqrt{\frac{4}{3}}D_{2}+\sqrt{\frac{1}{3}}D_{3}\right)=\frac{1}{2}\Big(Z_{1}\otimes \mathbb{1}_{2}\Big)\,.
    \end{split}
\end{equation}
Therefore it is possible for Gell-Mann letters to form a KAQ basis, so long as the metric has appropriate degeneracy patterns. These degeneracies are required to be able to translate Gell-Mann letters to Pauli words using more general maps than inner automorphisms. As an example consider $A_{1}$ and $A_{6}$. Their weights must be the same to construct the words $E_{1}$ and $E_{2}$. It is these considerations that lead us to the following UKAQ metrics 
\begin{equation}
\begin{split}
    \omega_{\rm UKAQ}(\vec{r},\vec{m})&=e^{ r_{1}}\bigg(|A_{1}\rangle\langle A_{1}|+|A_{6}\rangle\langle A_{6}|\bigg)+e^{r_{2}}\bigg(|A_{2}\rangle\langle A_{2}|+|A_{5}\rangle\langle A_{5}|\bigg) \\
    &+e^{ r_{3}}\bigg(|A_{3}\rangle\langle A_{3}|+|A_{4}\rangle\langle A_{4}|\bigg)+e^{m_{1}}\bigg(|S_{1}\rangle\langle S_{1}|+|S_{6}\rangle\langle S_{6}|\bigg) \\
    &+e^{m_{2}}\bigg(|S_{2}\rangle\langle S_{2}|+|S_{5}\rangle\langle S_{5}|\bigg)+e^{m_{3}}\bigg(|S_{3}\rangle\langle S_{3}|+|S_{4}\rangle\langle S_{4}|\bigg) \\
    &+e^{-\Delta}\bigg(|D_{1}\rangle\langle D_{1}|+|D_{2}\rangle\langle D_{2}|+|D_{3}\rangle\langle D_{3}|\bigg) \,,
\end{split}
\end{equation}
where we fix the determinant by setting $\Delta=\frac{2}{3}\sum_{i}(r_{i}+m_{i})$. Now we may see that the UKAQ parameterization is compatible with a many-body structure. A physically interesting structure arises when
\begin{equation}
    \begin{split}
        &r_{i}>0 \\
        &m_{i}<0 \\ 
        &\Delta \geq 0\,,
    \end{split}
\end{equation}
where a many-body structure exists if an inner automorphism $\Phi_{\rm MBP}$ exists mapping $\mathfrak{R}$ to the set of 1-string Pauli words. The inner automorphism for this example is constructed by the unitary 

\begin{equation}
    U_{\rm MBP}=\exp\left[\frac{i\pi}{4}Y_{1}\otimes Y_{2}\right]
\end{equation}
which maps computational states to Bell states. Therefore, with a few assumptions about the singular values, we have shown that many-body KAQ metrics exist which generalize the $\mathfrak{so}(4)$ Jensen metrics. For further exploration in the more complex loss functionals, we make the simplifying assumption that $m_{i}=-\Delta$ and $r_{2}=r_{3}$. This results in a class of ad$\mathfrak{R}_{0}$ invariant metrics 
\begin{equation}\label{eq:UKAQ_metric}
\begin{split}
  &\Omega_{\rm UKAQ}(t,s)=e^{\frac{t}{6}}\bigg(|A_{1}\rangle\langle A_{1}|+|A_{6}\rangle\langle A_{6}|\bigg)+e^{\frac{s}{6}}\Big(|A_{2}\rangle\langle A_{2}|+|A_{5}\rangle\langle A_{5}|+|A_{3}\rangle\langle A_{3}|+|A_{4}\rangle\langle A_{4}|\Big) \\
&+e^{-\frac{t+2s}{27}}\Big(|S_{1}\rangle\langle S_{1}|+|S_{6}\rangle\langle S_{6}|+|S_{2}\rangle\langle S_{2}|+|S_{5}\rangle\langle S_{5}|+|S_{3}\rangle\langle S_{3}|+|S_{4}\rangle\langle S_{4}|\\& +|D_{1}\rangle\langle D_{1}|+|D_{2}\rangle\langle D_{2}|+|D_{3}\rangle\langle D_{3}|\Big)
\end{split}
\end{equation}
where $\mathfrak{R}_{0}$=span$(A_{1},A_{6})$. Notice that  $\Omega_{\rm UKAQ}(t,t)$ corresponds to the class of $\mathfrak{so}(4)$ Jensen metrics, and all metrics off of this line are non-Jensen. We note this as over $\mathfrak{R}$ the metric contains two unequal weights whenever $t\neq s$.
\subsubsection{Partially translated KAQ metrics}
 The partially translated or PKAQ metrics have a KAQ basis which is mixed between Gell-Mann letters and Pauli words. They will be simpler to decode than UKAQ metrics, but still require some work. A natural class of PKAQ metric structures appear from a generalization of the $\mathfrak{sp}(2)$ Jensen metrics. The KAQ basis in this case must be compatible with the involution $\theta[X]=\mathcal{J}X^{\rm T}\mathcal{J}$ where the matrix $\mathcal{J}$ is defined as
\begin{equation}
\mathcal{J}=\begin{bmatrix}
        0 & \mathbb{1}_{2^{d-1}} \\
        -\mathbb{1}_{2^{d-1}} &0 
    \end{bmatrix}
\end{equation}
A critical difference appearing for this choice of $\theta$ is that most Gell-Mann matrices are not eigenvectors. For example
\begin{equation}
\begin{split}
    &\theta[A_{1}]=-A_{6} \\
    &\theta[A_{6}]=-A_{1} 
\end{split}
\end{equation}
thus $\big(A_{1}-A_{6}\big)$ is an eigenvector of $\theta$ that lives in $\mathfrak{R}$ and $\big(A_{1}+A_{6}\big)$ is an eigenvector of $\theta$ that lives in $\mathfrak{M}$. So we must use an orthonormal basis of the Killing-Cartan metric that is mixed between Gell-Mann letters and Pauli words to be compatible with $\theta$. Using the same dictionary, but a different Cartan decomposition, the Pauli words in $\mathfrak{R}$ are
\begin{equation}
    \begin{split}
        &(S_{1}-S_{6})=\frac{1}{2}\Big(Z_{1}\otimes X_{2}\Big), \quad  (A_{1}-A_{6})=\frac{1}{2}\Big(Z_{1}\otimes Y_{2}\Big)\\ 
         &(S_{2}+S_{5})=\frac{1}{2}\Big(X_{1}\otimes\mathbb{1}_{2}\Big), \quad (A_{2}+A_{5}) =\frac{1}{2}\Big(Y_{1}\otimes \mathbb{1}_{2} \Big)\\  &(S_{3}+S_{4})=\frac{1}{2}\Big(X_{1} \otimes X_{2}\Big), \quad (S_{4}-S_{3})=\frac{1}{2}\Big(Y_{1}\otimes Y_{2}\Big) \\  &(A_{3}+A_{4})=\frac{1}{2}\Big(Y_{1}\otimes X_{2}\Big), \quad  (A_{3}-A_{4})=\frac{1}{2}\Big(X_{1}\otimes Y_{2} \Big)\\  
         &D_{1}-\sqrt{\frac{1}{3}}D_{2}+\sqrt{\frac{2}{3}}D_{3}=\frac{1}{2}\Big(\mathbb{1}_{1}\otimes Z_{2}\Big) \\
         0&D_{1}+\sqrt{\frac{4}{3}}D_{2}+\sqrt{\frac{2}{3}}D_{3}=\frac{1}{2}\Big(Z_{1}\otimes \mathbb{1}_{2}\Big)\,,
    \end{split}
\end{equation}
and the Pauli words constructed from $\mathfrak{M}$ are
\begin{equation}
    \begin{split}
           &(S_{1}+S_{6})=\frac{1}{2}\Big(\mathbb{1}_{1}\otimes X_{2} \Big),\quad  (A_{1}+A_{6})=\frac{1}{2}\Big(\mathbb{1}_{1}\otimes Y_{2} \Big) \\ 
         &(S_{2}-S_{5})=\frac{1}{2}\Big(X_{1}\otimes Z_{2} \Big),\quad (A_{2}-A_{5})=\frac{1}{2}\Big(Y_{1}\otimes Z_{2}\Big) \\  
         &D_{1}+\sqrt{\frac{1}{3}}D_{2}-\sqrt{\frac{2}{3}}D_{3}=\frac{1}{2}\Big( Z_{1} \otimes Z_{2}\Big)\,.
    \end{split}
\end{equation}
Each word constructed from the same pair of Gell-Mann letters living in different Cartan blocks must be pre-translated into the Killing-Cartan basis to be compatible with $\theta$. 
 For each of these pre-translated words, we may assign an independent weight in the metric. Only the pairs $(S_{3},S_{4})$ and $(A_{3},A_{4})$ have the same weights, as we do not require them to be translated to achieve the Cartan decomposition. We are thus lead to the following parameterization for generalized $\mathfrak{sp}(2)$ metrics 
\begin{equation}
    \begin{split}
    \omega_{\rm PKAQ}(r_{i},m_{\alpha}) &= e^{r_{1}}|Z_{1}X_{2}\rangle \langle Z_{1}X_{2}|+e^{r_{2}}|Z_{1}Y_{2}\rangle \langle Z_{1}Y_{2}|+e^{r_{3}}|X_{1}\rangle \langle X_{1}|+e^{r_{4}}|Y_{1}\rangle \langle Y_{1}| \\
&+e^{r_{5}}\big(|S_{3}\rangle\langle S_{3}|+|S_{4}\rangle\langle S_{4}|\big)+e^{r_{6}}\big(|A_{3}\rangle\langle A_{3}|+|A_{4}\rangle\langle A_{4}|\big) \\
&+e^{r_{7}}|Z_{2}\rangle \langle Z_{2}|+e^{r_{8}}|Z_{1}\rangle \langle Z_{1}|+e^{m_{1}}|X_{2}\rangle \langle X_{2}|+e^{m_{2}}|Y_{2}\rangle \langle Y_{2}| \\ &+e^{m_{3}}|X_{1}Z_{2}\rangle \langle X_{1}Z_{2}|+e^{m_{4}}|Y_{1}Z_{2}\rangle \langle Y_{1}Z_{2}|+e^{-\Delta}|Z_{1}Z_{2}\rangle \langle Z_{1}Z_{2}| 
    \end{split}
\end{equation}
where again $\Delta$ is chosen such that ${\rm det}\big(\omega_{\rm PKAQ}\big)=1$. Here we clearly see that a few principle axes are still untranslated, but there is far less translation to be done than in the UKAQ example. Note as well that there are many other ways to construct a PKAQ parameterization that are not compatible with the Cartan involution. 

Again we take a simpler PKAQ parameterization for further investigation. We reduce to two parameters as this is the minimal number necessary to search for non-Jensen type critical points. We take the following non-naturally reductive ad$\mathfrak{R}_{0}$ invariant parameterization
\begin{equation}\label{eq:PKAQ metric}
\begin{split}
    \Omega_{\rm PKAQ}(t,s)=&e^{-\frac{3t+2s}{25}}\Big(|X_{2}\rangle \langle X_{2}|+e^{m_{2}}|Y_{2}\rangle \langle Y_{2}|+|X_{1}Z_{2}\rangle \langle X_{1}Z_{2}|+|Y_{1}Z_{2}\rangle \langle Y_{1}Z_{2}|+|Z_{1}Z_{2}\rangle \langle Z_{1}Z_{2}|\Big)  \\ +&e^{\frac{t}{10}}\Big(|Z_{1}X_{2}\rangle \langle Z_{1}X_{2}|+|Z_{1}Y_{2}\rangle \langle Z_{1}Y_{2}|+|X_{1}\rangle \langle X_{1}|+|Y_{1}\rangle \langle Y_{1}|+|Z_{2}\rangle \langle Z_{2}|+|Z_{1}\rangle \langle Z_{1}|\Big)  \\+&e^{\frac{s}{10}}\Big(|S_{3}\rangle\langle S_{3}|+|S_{4}\rangle\langle S_{4}|+|A_{3}\rangle\langle A_{3}|+|A_{4}\rangle\langle A_{4}|\Big) 
\end{split}
\end{equation}
which is bi-invariant with respect to $\mathfrak{R}_{0}$=span$(Z_{1},Z_{2})$. When $t=s$ the parametrization reduces to that of $\mathfrak{sp}(2)$ Jensen metrics, therefore any critical points found along the line $t=s$ in the $t-s$ plane are Jensen metrics for the given parameterization. Otherwise the other metrics in the $t-s$ plane are all non-Jensen. 
\subsubsection{Fully translated KAQ metrics}
\label{sec:FKAQdefine}
The final cipher class we consider is the class of fully translated KAQ (FKAQ) metrics. These metrics have no degeneracy, therefore to know about qubits the principal axes must already agree with a set of Pauli words. We use the following parameterization for FKAQ metrics

\begin{equation}
\begin{split}
    \omega_{\rm FKAQ}(w_{i},W_{\alpha})&=e^{w_{1}}|X_{1}\rangle\langle X_{1}|+e^{w_{2}}|Y_{1}\rangle\langle Y_{1}|+e^{w_{3}}|Z_{1}\rangle\langle Z_{1}|\\&+e^{w_{4}}|X_{2}\rangle\langle X_{2}|+e^{w_{5}}|Y_{2}\rangle\langle Y_{2}|+e^{w_{6}}|Z_{2}\rangle\langle Z_{2}| \\&+e^{W_{1}}|X_{1}X_{2}\rangle\langle X_{1}X_{2}|+e^{W_{2}}|X_{1}Y_{2}\rangle\langle X_{1}Y_{2}|+e^{W_{3}}|X_{1}Z_{2}\rangle\langle X_{1}Z_{2}| \\&+e^{W_{4}}|Y_{1}X_{2}\rangle\langle Y_{1}X_{2}|+e^{W_{5}}|Y_{1}Y_{2}\rangle\langle Y_{1}Y_{2}|+e^{W_{6}}|Y_{1}Z_{2}\rangle\langle Y_{1}Z_{2}|\\&+e^{W_{7}}|Z_{1}X_{2}\rangle\langle Z_{1}X_{2}|+e^{W_{8}}|Z_{1}Y_{2}\rangle\langle Z_{1}Y_{2}|+e^{-\Delta}|Z_{1}Z_{2}\rangle\langle Z_{1}Z_{2}|
\end{split}
\end{equation}
where we have assumed that all the principal axes are pure tensor. We have reduced the number of parameters by choosing a set of lab frames for the qubits, i.e. a choice of $xyz$-axes. But notice that this choice has no effect on the shape of the curvature functionals, changing the definition of axis or even meronomic frame comes only at the cost of performing an inner automorphism. The parameter $\Delta$ is chosen such that $\omega_{\rm FKAQ}$ has unit determinant. In this class no decoding needs to be done, only the commutation relations need to be checked to confirm the KAQ property.

Again we reduce the number of independent weights to simplify the search. First we consider a naturally reductive example based upon a penalty metric \cite{Nielsen:2005, Nielsen:2006, Brown_2018,brown2021effective}. We define the biased penalty metric 
\begin{equation}\label{eq:BPmetric}
\begin{split}
    \Omega_{\rm BP}(t,s)= e^{\frac{t}{6}}&\Big(|X_{1}\rangle\langle X_{1}|+|Y_{1}\rangle\langle Y_{1}|+|Z_{1}\rangle\langle Z_{1}|\Big)+e^{\frac{s}{6}}\Big(|X_{2}\rangle\langle X_{2}|+|Y_{2}\rangle\langle Y_{2}|+|Z_{2}\rangle\langle Z_{2}|\Big) \\+e^{-\frac{(t+s)}{18}}&\Bigg(|X_{1}X_{2}\rangle\langle X_{1}X_{2}|+|X_{1}Y_{2}\rangle\langle X_{1}Y_{2}|+|X_{1}Z_{2}\rangle\langle X_{1}Z_{2}| \\&+|Y_{1}X_{2}\rangle\langle Y_{1}X_{2}|+|Y_{1}Y_{2}\rangle\langle Y_{1}Y_{2}|+|Y_{1}Z_{2}\rangle\langle Y_{1}Z_{2}|\\ &+|Z_{1}X_{2}\rangle\langle Z_{1}X_{2}| +|Z_{1}Y_{2}\rangle\langle Z_{1}Y_{2}|+|Z_{1}Z_{2}\rangle\langle Z_{1}Z_{2}|\Bigg)\,.
\end{split}
\end{equation}
Motivated by the construction in \cite{Su:2006} we define a class of non-naturally reductive class of metrics, referred to as Abelian breakdown (AB)
\begin{equation}
\begin{split}
    \omega_{\rm AB}(t_{i})&=e^{-\Delta}\big(|Z_{1}\rangle\langle Z_{1}|+|Z_{2}\rangle\langle Z_{2}|+|Z_{1}Z_{2}\rangle\langle Z_{1}Z_{2}|\big) \\ &+e^{t_{1}}\Big(|X_{1}\rangle\langle X_{1}|+|X_{1}Z_{2}\rangle\langle X_{1}Z_{2}|\Big)+e^{t_{2}}\Big(|Y_{1}\rangle\langle Y_{1}|+|Y_{1}Z_{2}\rangle\langle Y_{1}Z_{2}|\Big)\\ &+e^{t_{3}}\Big(|X_{2}\rangle\langle X_{2}|+|Z_{1}X_{2}\rangle\langle Z_{1}X_{2}|\Big)+e^{t_{4}}\Big(|Y_{2}\rangle\langle Y_{2}|+|Z_{1}Y_{2}\rangle\langle Z_{1}Y_{2}|\Big)\\ &+e^{t_{5}}\Big(|X_{1}X_{2}\rangle\langle X_{1}X_{2}|+|Y_{1}Y_{2}\rangle\langle Y_{1}Y_{2}|\Big)+e^{t_{6}}\Big(|X_{1}Y_{2}\rangle\langle X_{1}Y_{2}|+|Y_{1}X_{2}\rangle\langle Y_{1}X_{2}|\Big)\,.
\end{split}
\end{equation}
The name is chosen as all degenerate eigenspaces form Abelian subalgebras. We may reduce to two parameters for further investigation
\begin{equation}
\label{eq:AB}
\begin{split}
    \Omega_{\rm AB}(t,s)&=e^{-\frac{8t+4s}{30}}\big(|Z_{1}\rangle\langle Z_{1}|+|Z_{2}\rangle\langle Z_{2}|+|Z_{1}Z_{2}\rangle\langle Z_{1}Z_{2}|\big) \\ &+e^{\frac{t}{10}}\Big(|X_{1}\rangle\langle X_{1}|+|Y_{1}\rangle\langle Y_{1}|+|X_{2}\rangle\langle X_{2}|+|Y_{2}\rangle\langle Y_{2}|\\& \;\;\;\;\;\;\;\;\;\;\;\;+|X_{1}Z_{2}\rangle\langle X_{1}Z_{2}|+|Y_{1}Z_{2}\rangle\langle Y_{1}Z_{2}|+|Z_{1}X_{2}\rangle\langle Z_{1}X_{2}|+|Z_{1}Y_{2}\rangle\langle Z_{1}Y_{2}|\Big)\\ &+e^{\frac{s}{10}}\Big(|X_{1}X_{2}\rangle\langle X_{1}X_{2}|+|Y_{1}X_{2}\rangle\langle Y_{1}X_{2}|+|X_{1}Y_{2}\rangle\langle X_{1}Y_{2}|+|Y_{1}Y_{2}\rangle\langle Y_{1}Y_{2}|\Big)\,.
\end{split}
\end{equation}
Notice that when $t=s$, a 1-dimensional space of naturally reductive metrics is obtained. But none of the contained naturally reductive metrics are Jensen metrics. To see this note that while the 3 dimensional subspace does form a subalgebra, the final commutation relation $[\mathfrak{M},\mathfrak{M}]\in \mathfrak{R}$ does not hold.  
\section{Equations of motion}
\label{sec:EquationsOfMotion}
We turn our attention now to the loss functionals we consider in this work, which  may only depend on the metric and structure constants. We consider loss functionals derived from curvature functionals  which are essentially tensor networks of the Christoffel connection. These classes of loss functionals contain structure constant networks not appearing in \cite{Freedman:2021a,Freedman:2021b,Freedman:2021c} at the same order in "perturbation" (defined by the number of structure constants in the contraction). We find it appealing to construct the loss functional from curvature functionals as \begin{enumerate}[I]
    \item $R[g]$ is the lowest order term in many natural classes of such actions. \item $R[g]$ always has two distinct classes of KAQ critical metrics.
\end{enumerate} 
A larger, natural class of loss functionals is
 \begin{equation}
 \label{eq:allCurveLoss}
 \begin{split}
     \mathcal{L}[g]&=R+\alpha R^{2}+\beta R_{ab}R^{ab}+\gamma R_{abcd}R^{abcd} \\
     &=R+\alpha\mathcal{R}_{0}+\beta \mathcal{R}_{2}+\gamma \mathcal{R}_{4}\, .
\end{split}
\end{equation}
 An additional improvement on the pure Ricci theory is that this loss functional contains order parameters, allowing for the appearance of quantum subsystems through spontaneous symmetry breaking. The results of \cite{Patterson,1977493} provide the equations of motion that must be satisfied in order for $g$ to be a critical metric of the loss functional $\mathcal{L}[g;\alpha,\beta,\gamma]$, over a compact manifold while enforcing a fixed volume element. The last point is why the determinant of the metric must be fixed for our purposes.

Performing the variation of these functionals (in an orthonormal basis of the metric) yields the following "stress-energy" tensors 
\begin{equation*}
\begin{split}
&T_{ij}=R_{ij}-\frac{1}{2}R\delta_{ij} \\
&T^{(0)}_{ij}=2RR_{ij}+2\nabla^{k}\nabla_{k}(R\delta_{ij})-2\nabla_{i}\nabla_{j}R-\frac{1}{2}R^{2}\delta_{ij}\\
&T^{(2)}_{ij}=2R_{ikjl}R^{kl}+\nabla^{k}\nabla_{k}R_{ij}+\frac{1}{2}\nabla^{k}\nabla_{k}(R\delta_{ij})-\nabla_{i}\nabla_{j}R-\frac{1}{2}R_{kl}R^{kl}\delta_{ij} \\&T^{(4)}_{ij}= 2R_{iklm}R\indices{_j^{klm}}+4R_{ikjl}R^{kl}+4\nabla^{k}\nabla_{k}R_{ij}-2\nabla_{i}\nabla_{j}R-4R_{ik}R\indices{^k_j}-\frac{1}{2}R_{klmn}R^{klmn}\delta_{ij}
\end{split}   \,. 
\end{equation*}
It is important to note that using our definition of $R_{ijkl}$ (see Appendix A), leads to different indicies being contracted in the Riemann-Ricci terms than those found in \cite{Patterson}. Muto \cite{Muto:1977} proved that a given metric is a critical point iff
\begin{equation}
    \mathcal{T}_{ij}=T_{ij}+\alpha T^{(0)}_{ij}+\beta T^{(2)}_{ij}+\gamma T^{(4)}_{ij}=\Lambda \delta_{ij}
\end{equation}
where $\Lambda$ is a real constant depending on the parameters of the problem, namely $m$, $r$, and the coupling constants. Note to reduce the calculation, we may move all the terms already proportional to the metric to the RHS. Further, as we are considering homogeneous spaces the covariant derivatives of the Ricci scalar vanish. The "stress-energy" tensors simplify to 
\begin{equation*}
\begin{split}
&T_{ij}=R_{ij} \\
&T^{(0)}_{ij}=2RR_{ij}+2R\nabla_{k}\nabla_{k}(\delta_{ij})\\
&T^{(2)}_{ij}=2R_{ikjl}R_{kl}+\nabla_{k}\nabla_{k}(R_{ij})+\frac{1}{2}R\nabla_{k}\nabla_{k}(\delta_{ij}) \\
       &T^{(4)}_{ij}= 2R_{iklm}R_{jklm}+4R_{ikjl}R_{kl}+4\nabla_{k}\nabla_{k}(R_{ij})-4R_{ik}R_{kj},
\end{split}    
\end{equation*}
where, as we are working in an orthonormal basis, we are free to lower all the indices. Summation over repeated indices is still implied. The last simplification available is to compute the terms involving the covariant Laplacian. The Laplacian of the metric vanishes
\begin{equation} 
    \begin{split}
        \nabla_{k}\nabla_{k}(\delta_{ij})&= -\nabla_{k}(\Gamma_{lki}\delta_{lj}+\Gamma_{lkj}\delta_{il})= \nabla_{k}(\Gamma_{ikj}+\Gamma_{jki}) =0
    \end{split}
\end{equation}
which is equivalent to metric compatibility of the connection. We need to also compute the Laplacian of the Ricci tensor, but notice if we take a particular choice of the coupling constants the contribution from this term cancel. This particular combination is in fact the Gauss-Bonnet term yielding the loss functional 
\begin{equation}
\label{eq:GBloss}
     \mathcal{L}_{\rm GB}(\gamma)=R+\gamma(\mathcal{R}_{0}-4\mathcal{R}_{2}+\mathcal{R}_{4}) \,.
 \end{equation}
We should emphasis that the Gauss-Bonnet term is not topological in this theory; the dimension of the space is $4^{d}-1\geq 15$, thus clearly never 4. The equations of motion for the chosen loss functional are
\begin{equation}\label{eq:GBeom}
    \mathcal{T}_{ij}=R_{ij}+\gamma\Big(2RR_{ij}+2R_{iklm}R_{jklm}-4R_{iljk}R_{lk}-4R_{ik}R_{kj}\Big)=\Lambda_{\rm GB} \delta_{ij}\,.
\end{equation}
where indeed we see the $\nabla^{2}R_{ij}$ term vanishes. In the next section we provide much evidence that the Gauss-Bonnet functional has the Killing-Cartan geometry as an unstable critical point. We suspect that the general class of loss functionals contained in Eq.\ref{eq:allCurveLoss} are concave about the Killing-Cartan geometry for the same reason as found in \cite{Freedman:2021a}, who examined the behavior of individual diagrams contributing to the functionals (see their Appendix C and our Appendix \ref{app:networks}). 

And while ostensibly we have made a restrictive choice of coupling constants, by using a graphical method developed in Appendix \ref{app:networks} we in fact see that the loss functional in Eq.(\ref{eq:GBloss}) has general properties of the larger family, as no special cancellations appear in the graphs. As further evidence we may also contrast the Gauss-Bonnet loss functional with those introduced by Freedman and Zini. Besides the Ricci scalar, they considered non-Gaussian functionals for example the Euclidean integral
\begin{equation}
\label{eq:FZNGloss}
F[G;\kappa]=\int_{\mathbb{R}^{3(4^{n}-1})}dy_{1}dy_{2}dy_{3}e^{\left[-\kappa\sum_{i=1
}^{3}g_{ab}y^{a}_{i}y^{b}_{i}+C_{abc}y^{a}_{1}y^{b}_{2}y^{c}_{3}\right]} \,.
\end{equation}
The use of three integration variables is necessary to construct a non-vanishing scalar from the structure constants. Considering perturbed Gaussian integrals allows for a systematic approach to the perturbative calculation.
 In this way a series of trivalent tensor networks is obtained that determines the loss functional at a given order in perturbation parameter $\kappa$.

Now consider the graphical representation of the types of terms is given in Appendix \ref{app:networks}. The diagrams help illuminate a few important points of comparison between the loss functionals in Eq.~(\ref{eq:allCurveLoss}) compared to Eq.~(\ref{eq:FZNGloss}). First, the family of curvature terms depends on diagrams with only at most four structure constants in the contraction. In contrast, the perturbed Gaussian of \cite{Freedman:2021a} contains a series out to infinite order, which was computed up to terms of order six for the analysis. At the level of fourth order terms, the Gauss-Bonnet combination does not induce any special cancellation between diagrams appearing in $\mathcal{R}_{0}$, $\mathcal{R}_{2}$, and $\mathcal{R}_{4}$. The individual curvature terms contain somewhat symmetric combinations of diagrams with a different relationship from that imposed by Eq.~(\ref{eq:FZNGloss}). This family of loss-functionals is well-suited to a geometrically illuminating study that may be able to connect KAQ structure to other known and interesting classes of metrics, including the naturally reductive metrics.

\section{Results for SU(4)}
\label{sec:results}
The equations of motion allow us to search for solutions in the space of each 2-parameter KAQ metrics defined in Section \ref{sec:KAQparam}. These parameterizations include both naturally and non-naturally reductive metrics. To yield a solution, a given $\omega$ must generate a diagonal stress energy tensor satisfying  
\begin{equation}
    \mathcal{T}_{i}\equiv\mathcal{T}_{ii}=\Lambda_{\rm GB}
\end{equation}
where $1\leq i\leq15$ and $\Lambda_{\rm GB}$ is real. Thus for a given parameterization, we first determine the number of independent $\mathcal{T}_{i}$. By setting $s=at$, we simultaneously plot the independent $\mathcal{T}_{i}$, allowing us to vary $a$ and check if any critical points appear for non-zero values of $t$. 

By making contour plots of the loss function, we  systematically find potential critical points and check if they are indeed true critical points. Further, while we do not have access to the second variation of the loss functional, we still obtain information about the second derivative by comparing contour plots which agree along the line $t=s$. Doing so affords us a glimpse at the stability of the loss functional around certain Jensen type critical points. The straightforward but lengthy evaluation of the second variation could be performed to fully check stability.

\subsection{FKAQ critical points}
Here we collect our results for the two FKAQ parameterizations given in Section \ref{sec:KAQparam}. These include the naturally reductive parameterization $\Omega_{\rm BP}$ defined in Eq.(\ref{eq:BPmetric}) and the non-naturally reductive parameterization $\Omega_{\rm AB}$ defined in Eq.(\ref{eq:AB}). Using two parameterizations allows us to compare the type of critical metrics that appear in the naturally reductive vs. non-naturally reductive cases.  

We show results with the help of two types of figures. First, to demonstrate the technique, we plot the equations of motion for an example, at fixed values of the weighting parameters in the metric ($s, t$). But, to visualize the space of solutions in the weighting parameters, we show contour plots of the loss functionals for varying $s, t$. It is important to stress that not all critical points that appear in these plots are critical in the space of left-invariant metrics. All that can be learned from these plots is whether they are critical in the considered parameterization space. In order to determine true criticality we always appeal to the equations of motion. 
\subsubsection{Biased penalty metric}
There are only three independent $\mathcal{T}_{i}$ for the biased penalty metric ansatz, $\Omega_{\rm BP}$ from Eq.(\ref{eq:BPmetric}). The equations that must be satisfied are 
\begin{align*}
 \mathcal{T}_{1}=e^{\frac{1}{9} (-7 a-10) t} \Big(-&3 \gamma  e^{\frac{a t}{3}}+18.75 \gamma  e^{\frac{4}{9} (a+1) t}+9\gamma 
   e^{\frac{2}{3} (a+2) t}-1.125 \gamma  e^{\frac{1}{9} (2 a+11) t}+1.125 \gamma  e^{\frac{1}{9} (8 a+5) t} \\ +&0.375 \gamma 
   e^{\frac{1}{9} (10 a+13) t}+0.75 e^{\frac{5}{9} (a+1) t}+0.25 e^{\frac{1}{9} (7 a+13) t}-1.125 \gamma  e^{t/3}\Big) \\ =&\Lambda_{\rm GB},
\end{align*}
\begin{align*}
 \mathcal{T}_{4}= e^{\frac{1}{9} (-10 a-7) t} \Big(-&1.125 \gamma  e^{\frac{a t}{3}}+18.75 \gamma  e^{\frac{4}{9} (a+1) t}+1.125 \gamma 
   e^{\frac{1}{9} (5 a+8) t}-1.125 \gamma  e^{\frac{1}{9} (11 a+2) t}\\+&0.375 \gamma  e^{\frac{1}{9} (13 a+10) t} +9 \gamma 
   e^{\frac{2}{3} (2 a t+t)}+0.75 e^{\frac{5}{9} (a+1) t}+0.25 e^{\frac{1}{9} (13 a+7) t}\\-&3 \gamma  e^{t/3}\Big) =\Lambda_{\rm GB},
\end{align*}
\begin{align}
\label{eq:EOMBP}
 \mathcal{T}_{7}= e^{-\frac{10}{9} (a+1) t} \Big(&2\gamma  e^{\frac{2 a t}{3}}+1.5 \gamma  e^{\frac{1}{3} (a+1) t}+51.5 \gamma  e^{\frac{8}{9}
   (a+1) t}+3 \gamma  e^{\left(a+\frac{4}{3}\right) t}-18.75 \gamma  e^{\frac{1}{9} (4 a+7) t}\nonumber\\
   -&0.75 \gamma  e^{\frac{1}{9} (5a+11) t}-18.75 \gamma  e^{\frac{1}{9} (7 a+4) t}-0.75 \gamma  e^{\frac{1}{9} (11 a+5) t} +3 \gamma  e^{\frac{4 a t}{3}+t}\nonumber\\
   -&0.5 e^{\frac{1}{9} (5 a+8) t}-0.5 e^{\frac{1}{9} (8 a+5) t}+2 e^{a t+t}+2 \gamma  e^{2 t/3}\Big) 
   \nonumber\\=&\Lambda_{\rm GB}.
\end{align}
With only three independent diagonal elements, this parameterization is only slightly more complicated than the Jensen type. For Jensen metrics there are only 2 independent $\mathcal{T}_{i}$. Recall that we want to find non-Jensen type critical points, as for larger number of qubits Jensen metrics are highly non-local and do not give a typical many-body structure. 

\begin{figure}[H]
\centering
\begin{subfigure}{.45\linewidth}
  \centering
\includegraphics[width=0.95\linewidth]{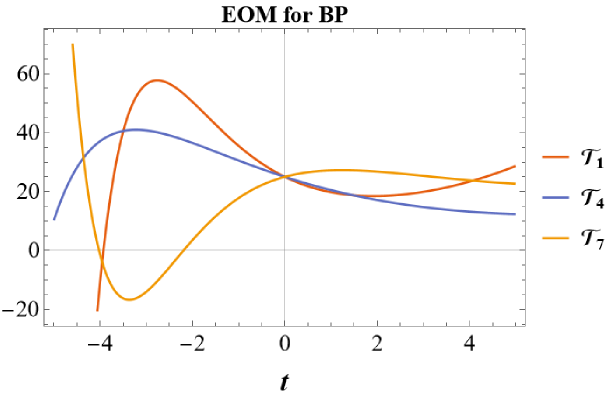}\label{fig:FKAQ_EOMJensen} 
\caption{$a=0.1$}
\end{subfigure} 
\begin{subfigure}{.45\linewidth}
  \centering
\includegraphics[width=0.95\linewidth]{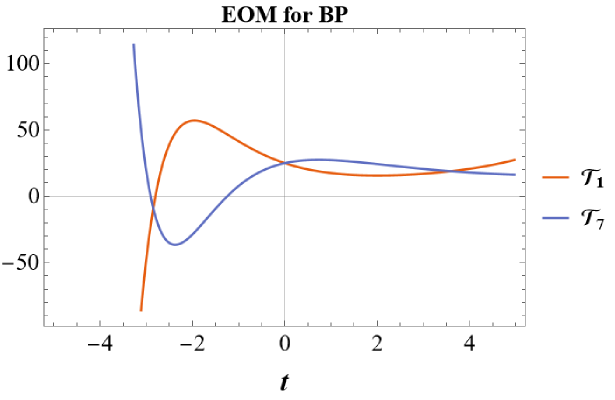}\label{fig:FKAQ_EOM1} 
 
\caption{$a=1$}
\end{subfigure}
\caption{Here the equations of motion (Eq.\ref{eq:GBloss})  are plotted for the biased penalty metric ansatz $\Omega_{\rm BP}$ (Eq.\ref{eq:BPmetric}), which includes a Gauss-Bonnet with $\gamma=1$. The left plot shows the line $t=s$, and for the biased penalty metric these are Jensen-type. In fact we find an additional solution (relative to pure Ricci scalar) for $t<0$. The second plot demonstrates that for most $a$ strictly left-invariant solutions do not exist.}
\label{fig:FKAQEOM1}
\end{figure}

\begin{figure}[H]
\centering
\includegraphics[width=0.45\linewidth]{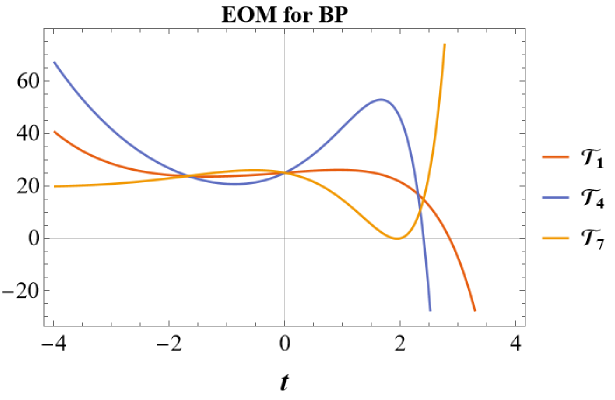}  
\label{fig:FKAQ_EOM_NonJensenCriticalPt}
\caption{Here the equations of motion for $\Omega_{\rm BP}$ are plotted for $a\approx -2.06$. We see that there is a non-Jensen type solution for $t\approx-1.67$. }
\label{fig:FKAQNJ}
\end{figure}

In Figure \ref{fig:FKAQEOM1}  we plot the set of $\mathcal{T}_{i}$ for two different values of $a$. The left plot shows that for most values of $a$ (here 0.1), the only solution that appears is the trivial non-KAQ solution $s=at=0$. But for special values of $a$, non-trivial solutions do appear. We unsurprisingly find Jensen type critical points for $a=1$, which is expected due to the simplicity of the equations of motion. Figure \ref{fig:FKAQNJ} however, shows an example of a non-Jensen type critical points, of which several exist. 
 
\begin{figure}[H]
\centering
\begin{subfigure}{.45\linewidth}
  \centering
\includegraphics[width=0.9\linewidth]{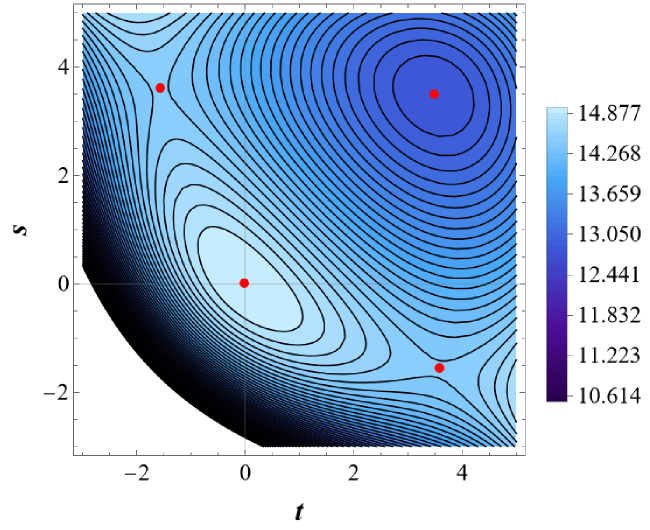}  \label{fig:FKAQ}
\caption{$R$}
\end{subfigure} 
\begin{subfigure}{.45\linewidth}
  \centering
\includegraphics[width=0.9\linewidth]{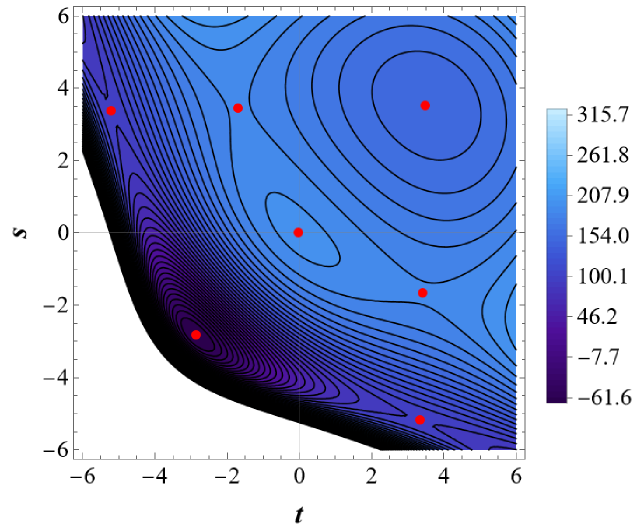}  
 \caption{$R+(\mathcal{R}_{0}-4\mathcal{R}_{2}+\mathcal{R}_{4})$}
\end{subfigure}
\caption{Contours of the loss functional are plotted for $\Omega_{\rm BP}$, Eq.(\ref{eq:BPmetric}), with $\gamma=0$ (the Ricci scalar only) and $\gamma=1$. All red points marked on the plots are critical in the space of left-invariant metrics with fixed determinant. Critical points residing on the line $t=s$ are Jensen type, while all others are non-Jensen appealing to the form of $\Omega_{\rm BP}(t,s)$.}
\label{fig:FKAQContours}
\end{figure}

In Fig \ref{fig:FKAQContours} we present a holistic view of the critical points that appear through the use of contour plots of the loss functional. We see that even for the Ricci scalar, non-Jensen naturally reductive critical points are present. The shapes of the contour near the origin show why evolutionary searches performed in \cite{Freedman:2021a} likely missed out on the more structured critical points. Evolutionary searches begin with the choice of an initial parent point. The natural choice here is the Killing-Cartan metric, or a randomly selected nearby point. The search then casts a small net around the parent point, and the loss functions is computed for each of these points. The point with the lowest value serves as the new parent point in the following step. The topography of these plots demonstrates that such a method can fail to find the saddle points, instead settling on the Jensen-type critical point(s).  

\subsubsection{Abelian breakdown metric}
\begin{figure}[H]
\centering
\begin{subfigure}{.45\linewidth}
  \centering
\includegraphics[width=0.95\linewidth]{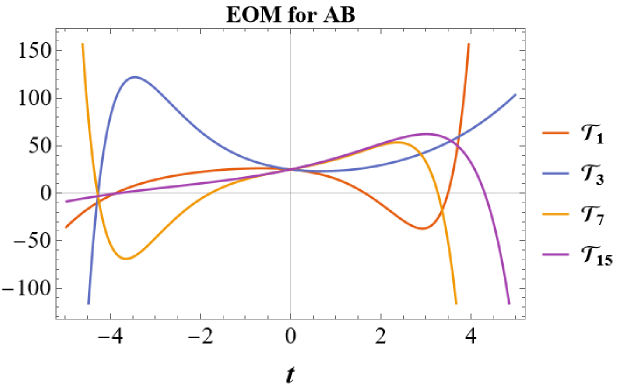}
\caption{$a=-2.06$}
\end{subfigure} 
\begin{subfigure}{.45\linewidth}
  \centering
\includegraphics[width=0.95\linewidth]{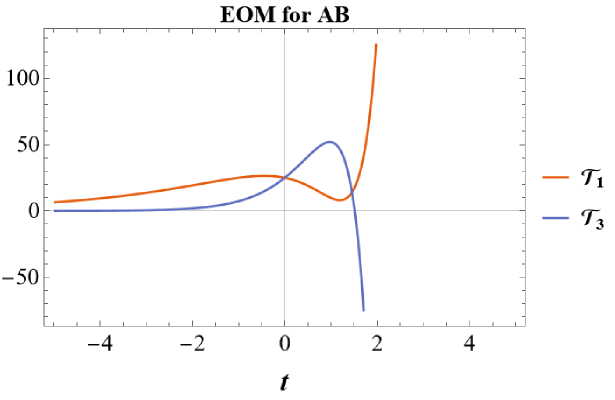} 
 \caption{$a=1$}
\end{subfigure}
\caption{Plotted are the equations of motion for the Abelian breakdown metric ansatz, $\Omega_{\rm AB}$ of Eq.(\ref{eq:AB}), which is non-naturally reductive. Again, we have taken $\gamma=1$ for the amplitude of the Gauss-Bonnet term. The first plot shows the behavior for a typical $a$; no non-trivial solutions exist. The second plot shows the existence of a non-Jensen type critical point for $a=1$ and $t\approx 1.45$.}
\label{fig:FKAQEOM4}
\end{figure}

Moving to the non-naturally reductive FKAQ parameterization, $\Omega_{\rm AB}$ from Eq.(\ref{eq:AB}), we plot the set of independent $\mathcal{T}_{i}$ in Fig \ref{fig:FKAQEOM4}, where there are 4 such elements. We find solutions only for $a=1$, the potential meeting point in the left plot never actually becomes a crossing. In this case the solution is naturally reductive but not of Jensen type, even though the corresponding critical metric only has two distinct weights, as the weights do not distinguish blocks that belong to a Cartan decomposition.  

\begin{figure}[H]
\centering
\begin{subfigure}{.45\linewidth}
  \centering
\includegraphics[width=0.9\linewidth]{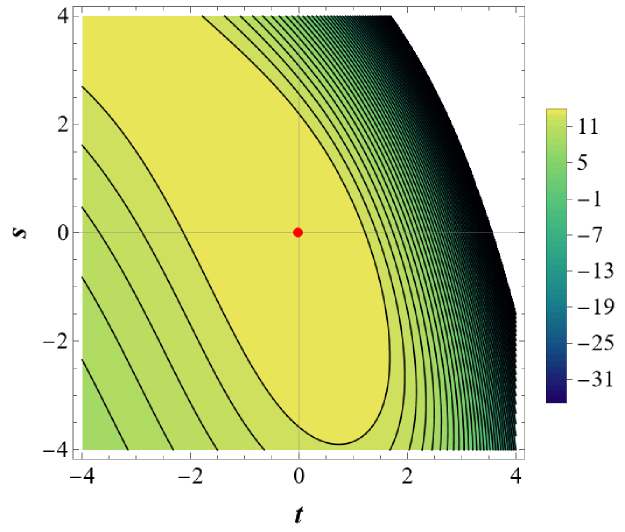}  \label{fig:FKAQ_Nonreductive1}
\caption{$R$}
\end{subfigure} 
\begin{subfigure}{.45\linewidth}
  \centering
\includegraphics[width=0.9\linewidth]{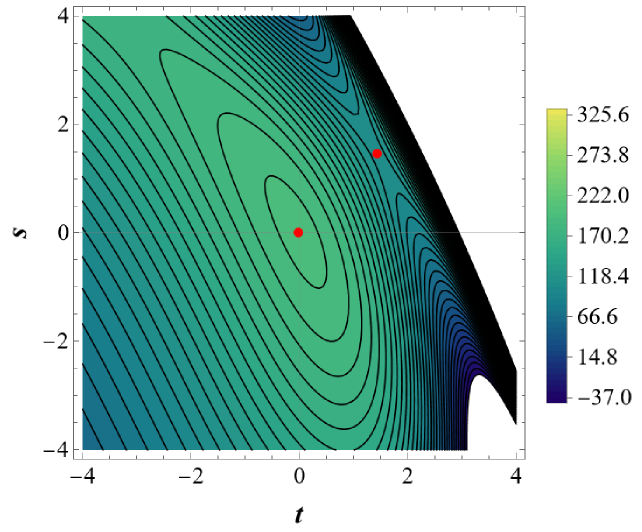}  
 \label{fig:FKAQ_Nonreductive2}
\caption{$R+(\mathcal{R}_{0}-4\mathcal{R}_{2}+\mathcal{R}_{4})$}
\end{subfigure}
\caption{Contours of the loss functional $\mathcal{L}_{\rm GB}$ is plotted for $\Omega_{\rm AB}$, Eq.(\ref{eq:AB}), with $\gamma=0$ (the Ricci scalar only) and $\gamma=1$. The red points marked on the plots are critical points in the space of left-invariant metrics with fixed determinant. As explained in Section \ref{sec:FKAQdefine}, no metric in this class, regardless of parameter values, is of the Jensen type.}
\label{fig:FKAQContoursNonreductive}
\end{figure}

We have plotted contour plots in Fig \ref{fig:FKAQContoursNonreductive}, where we find only one non-trivial solution. Comparing to the biased penalty example we may make two comments. First, there are far fewer critical points along these directions in the space of left-invariant metrics. Further, the critical point we obtained is rather interesting. The KAQ critical metric puts distinct weight on Cartan subalgebra distinguishing it from the remaining observables. While a small step, it is evidence that curvature-based loss functionals may support KAQ critical points with many-body local structure.  

\subsection{UKAQ critical points}
Turning to our UKAQ parametrization, $\Omega_{\rm UKAQ}$ from Eq.(\ref{eq:UKAQ_metric}), we may perform the same search for critical points. The equations of motions are much more complex for this example, where there are 6 independent $\mathcal{T}_{i}$. While certain choices of $a$ reduce the number of independent $\mathcal{T}_{i}$, only for $a=1$ do we obtain non-trivial solutions to the equations of motion. 

\begin{figure}[H]
\centering
\begin{subfigure}{.45\linewidth}
  \centering
\includegraphics[width=0.9\linewidth]{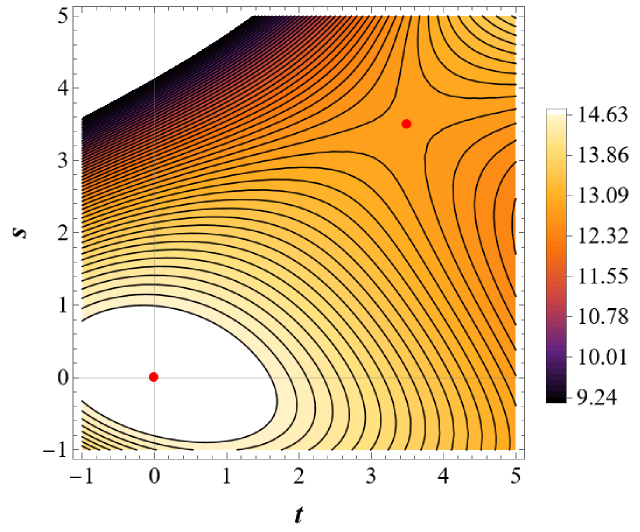}  
\caption{$R$}
\end{subfigure} 
\begin{subfigure}{.45\linewidth}
  \centering
\includegraphics[width=0.9\linewidth]{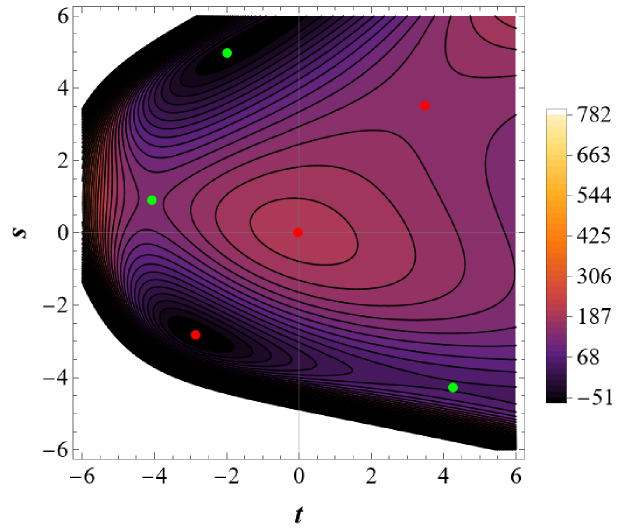}  
 \label{fig:UKAQ}
\caption{$R+(\mathcal{R}_{0}-4\mathcal{R}_{2}+\mathcal{R}_{4})$}
\end{subfigure}
\caption{Contours of the loss functionals are plotted for the $\Omega_{\rm UKAQ}$ parameterization, Eq.(\ref{eq:UKAQ_metric}). The red points located along the line $t=s$ are critical points in the space of left-invariant metrics. All other points (green) are only critical in the considered parameterization space.} 
\label{fig:UKAQcontours}
\end{figure}

By comparing Figure \ref{fig:FKAQContours} and Figure \ref{fig:UKAQcontours} we obtain some evidence about the preference of naturally reductive metric. Notice that these figures exactly agree on the line $t=s$, so we can see how the value of the functional changes in naturally reductive vs. non-naturally reductive direction. We see that Figure \ref{fig:FKAQContours} clearly contains more true critical metrics, and that the only critical metrics contained within Figure \ref{fig:UKAQcontours} are naturally reductive as well. This provides some evidence that naturally reductive metrics may be favored over non-naturally reductive metrics when weighed by curvature based functionals. 

Some understanding about the stability of the Jensen type critical points common to both examples can be obtained by comparing these figures. 
 The Jensen type critical point with $s,t<0$ appears to be a stable critical point, where as the solution for positive $t$ and $s$ is clearly a saddle point. Although there are many different cross sections one must consider to truly determine the stability.

We do not fully understand the propensity towards Jensen metrics in our examples. It could simply be a preference for naturally reductive metrics. There is however another possibility. At such a low dimension (two qubits) the Jensen metrics and penalty metrics are essentially the same. So, the loss functions investigated here may simply be favoring penalty structures. It would be enlightening to study $\mathfrak{su}(8)$ to illuminate this point.
\subsection{PKAQ critical points}
For the final parameterization, we investigate $\Omega_{\rm PKAQ}$ defined in Eq.(\ref{eq:PKAQ metric}). We only find Jensen-type critical points, and note that this direction seems the least fruitful in the search for critical metrics.

\begin{figure}[H]
\centering
\begin{subfigure}{.45\linewidth}
  \centering
\includegraphics[width=0.9\linewidth]{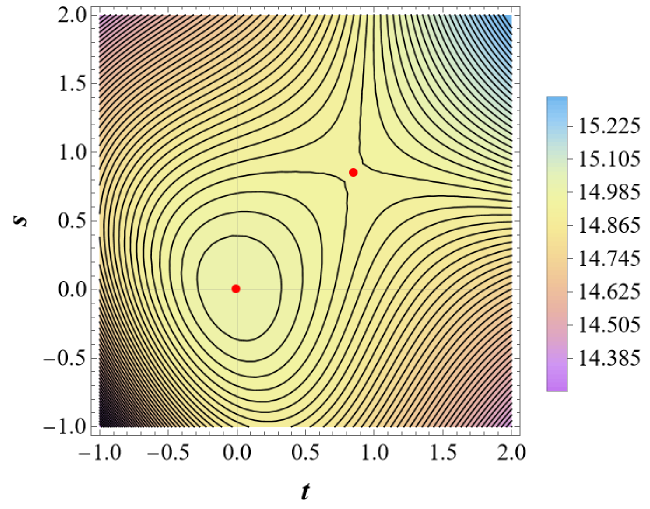}  
\caption{$R$}
\end{subfigure} 
\begin{subfigure}{.45\linewidth}
  \centering
\includegraphics[width=0.9\linewidth]{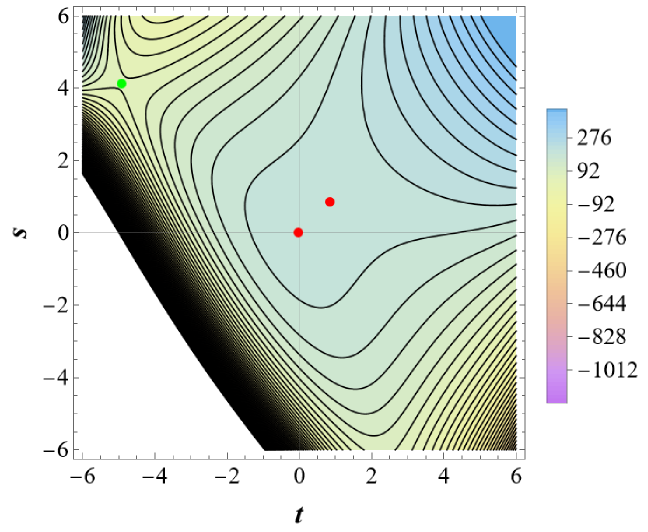}  
 \label{fig:PKAQ}
\caption{$R+(\mathcal{R}_{0}-4\mathcal{R}_{2}+\mathcal{R}_{4})$}
\end{subfigure}
\caption{
Contours are plotted for the $\Omega_{\rm PKAQ}$ parameterization, Eq.(\ref{eq:PKAQ metric}). The red points indicate true critical points, which are all Jensen type. The green point, in the second plot, is only critical in the parameterization space.}
\label{fig:fig}
\end{figure}

\section{Conclusions} 
\label{sec:Conclude}
In this work we have found critical points of curvature-based loss functionals that are KAQ and include metrics which are compatible with many-body physics. All of the critical points we have found are naturally reductive, although we cannot rule out the existence of non-naturally reductive KAQ metrics. From the examples we tried we found no ad$\mathfrak{R}_{0}$ invariant critical metrics which were not Jensen. We also have found that most critical points are saddle points, with the potential for a few stable Jensen critical points.

While we have only analyzed detailed examples for $\mathfrak{su}(4)$, it is straightforward, although numerically intensive, to generalize these constructions to larger $N$. What we have presented is evidence that the KAQ ansatz is a useful tool both for finding critical metrics among the large set of left-invariant metrics and for better understanding the structure of KAQ critical points. Even going to $\mathfrak{su}(8)$ would be helpful in determining more about the structure of the critical points of curvature-based loss functionals. For example with three qubits, penalty metrics are not Jensen, allowing for more exploration in what exactly drives the value of the loss function; the algebraic properties of the principal axes or structure of the weights?

These constructions for KAQ metrics we provide may also be used to explore a much broader family of functionals depending on curvature tensors. However, as this construction cannot find non-KAQ minima, we cannot study the relative frequency of KAQ vs non-KAQ. Ideally, one would like functionals with only KAQ minima, or related structure for $N\neq2^{d}$.



\section{Acknowledgements} 
We thank Sarang Gopalakrishnan for bringing the Freedman and Zini papers to our attention and for initial discussions. The work of S. Prudhoe and S. Shandera was supported by NSF PHY-2310662. The work of R. Kumar was supported by NSF PHY-2207851.
\bibliographystyle{unsrt}
\bibliography{dynmap,kaq}
\appendix 

\section{Curvature functionals from the structure constants} 
\label{app:functionals}

 To begin, a non-holonomic basis is a choice of basis over the tangent manifold that is non-commutative. Assume that $\{X_{a}\}$ forms a basis for the tangent manifold with
\begin{equation}
[X_{a},X_{b}]=C\indices{^c_{ab}}X_{c}
\end{equation}
defining the structure constants associated to the chosen basis. The Christoffel connection (the unique connection that is torsion-less and compatible with the metric) in a non-coordinate basis is
\begin{equation}
\Gamma_{abc}=\frac{1}{2}\left[\partial_{c}g_{ab}+\partial_{b}g_{ac}-\partial_{a}g_{bc}+C_{abc}-C_{bca}-C_{cba}\right]\,.
\end{equation}
Assuming $\{X_{a}\}$ is an orthonormal basis of the metric and that the metric is left invariant, the terms involving the metric vanish leaving  
\begin{equation}
\Gamma_{abc}=\frac{1}{2}(C_{abc}-C_{bca}-C_{cba})\,.
\end{equation}
Notice that $\Gamma_{abc}=-\Gamma_{cba}$. The components of the Christoffel connection in an orthonormal, non-holonomic basis are referred to as the Ricci rotation coefficients \cite{Wald:106274}. The Riemann tensor and its various traces are computed as functions of the Christoffel connection and structure constants \cite{https://doi.org/10.48550/arxiv.1711.09503}
\begin{equation}
\begin{split}
R_{abcd}=&\Gamma\indices{^e_{db}}\Gamma_{ace}-\Gamma\indices{^e_{cb}}\Gamma_{ade}-C\indices{^e_{cd}}\Gamma\indices{_{aeb}}
\\ R_{ab}=&-\frac{1}{2}g_{eb}\bigg(C_{dac}C^{ced}+C_{cad}C^{ced}-\frac{1}{2}C_{acd}C^{ecd}\bigg)
\\ R=&-\frac{1}{2}C_{abc}C^{cba}-\frac{1}{4}C_{abc}C^{abc} \,.
\end{split}
\end{equation}
Using the graphical method described in Appendix \ref{app:networks}, we compute the second order curvatures functionals in terms of structure constant networks. We have found 

\begin{equation*}
\begin{split}
\mathcal{R}_{2}[g]
=\frac{1}{16}g^{ef}g_{lk}\Bigg(&C_{dbf}\left[4C_{ace}C^{dbk}C^{acl}-4C_{ace}C^{bdk}C^{cla}-8C_{ace}C^{dbk}C^{cla}\right] \\ +&C_{fdb}\left[C_{eca}C^{kdb}C^{lca}-4C_{ace}C^{kdb}C^{acl}+4C_{ace}C^{kdb}C^{cla}\right]\Bigg)
\end{split}
\end{equation*}
\begin{equation*}
 \begin{split}   \mathcal{R}_{4}[g]=\frac{1}{8}g^{ef}g_{lk}\Bigg(&C_{dbf}\bigg[4C_{ace}C^{dbk}C^{acl} -4C_{ace}C^{bdk}C^{cla}-8C_{ace}C^{dbk}C^{cla}\bigg]\\+&C_{fdb}\bigg[3C_{eca}C^{kdb}C^{lca}+8C_{ace}C^{kdb}C^{acl}+8C_{ace}C^{kdb}C^{lca}\bigg] \\\\ -&C_{fcd}\bigg[C_{eba}C^{kdb}C^{lca}+2C_{aeb}C^{dbk}C^{lca}+4C_{aeb}C^{bdk}C^{acl} \\
&8C_{aeb}C^{kdb}C^{lca}-8C_{aeb}C^{kdb}C^{cla}+28C_{aeb}C^{kdb}C^{acl}\bigg] \Bigg)\,.
 \end{split}
\end{equation*}

\section{Graphical representation of structure networks}
\label{app:networks}

\usetikzlibrary{positioning, arrows.meta, decorations.markings, bending}
\tikzset{ 
  put token/.style={
          decoration={markings, mark=at position 0.4 with {\arrow{Stealth[length=8pt,open,bend,sep]}}, mark=at position 0.6 with {\arrow{Stealth[length=8pt,open,bend,sep]}},mark=at position 0.8 with {\arrow{Stealth[length=8pt,open,bend,sep]}}},
  },
}

The fundamental information about the group structure and the metric on the manifold is carried in the structure constants, organized into the connection, and the metric. Any functionals one might want to define, including the Ricci scalar $R$, $\mathcal{R}_{2}[g]$, and $\mathcal{R}_{4}[g]$ can be built from the totally contracted combinations of $C_{ijk}$, $C^{ijk}$, $g^{\ell m}$, and $g_{\ell m}$. A diagrammatic representation of the terms is useful, where each $C_{ijk}$ is represented by a node and contractions are indicated with edges. The line style of the edge carries relevant information about the contraction, i.e., whether it involves the metric and any symmetry information.

We define the following notation for the placement and style of nodes and edges: $C_{ijk}$ is a node with edges emerging upward, while $C^{ijk}$ is a node with edges emerging downward. Contractions via the metric will then always be represented by horizontal lines, while others appear as vertical or diagonal lines. Since the $C_{ijk}$ are antisymmetric in the last two indices, we differentiate contractions between indices in different positions. Contractions of indices both in the first position are drawn with a solid line, both indices in the second or both in the third position with a dashed line, and contractions between an index in the first position and one in the second or third is a solid line with arrows pointing from the node containing the first-position index to the node containing the second- or third-position index.

For example, the two terms contributing to the $R=-\frac{1}{2}C_{abc}C^{cba}-\frac{1}{4}C_{abc}C^{abc}$ can be drawn as
\begin{center}
\begin{tikzpicture}
[roundnode/.style={circle, draw=black!40, very thick, minimum size=8mm}]
    \node[roundnode,fill=cyan!20] (x1) {};
    \node[roundnode,fill=cyan!20] (x2) [below=1.5cm of x1] {};
     
    \draw[dashed, very thick] (x1) -- (x2) {};
    \draw[-, very thick] (x1.300) edge[put token, postaction={decorate}]  (x2.60) {};
    \draw[-, very thick] (x2.120) edge[put token, postaction={decorate}] (x1.240);

    \node[below] at (x2.south) {\textbf{$C_{abc}C^{cba}$}};

    \node[roundnode,fill=cyan!20] (y1) [right=2.5cm of x1] {};
    \node[roundnode,fill=cyan!20] (y2) [below=1.5cm of y1] {};
     
    \draw[dashed, very thick] (y1) -- (y2) {};
    \draw[dashed, very thick] (y1.300) -- (y2.60) {};
    \draw[-, very thick] (y2.120) -- (y1.240);

    \node[below]  at (y2.270) {\textbf{$C_{abc}C^{abc}$}};
    
\end{tikzpicture}
\end{center}
These are similar to the theta diagrams defined in \cite{Freedman:2021a}. The remaining curvature structures we consider, $\mathcal{R}_2$ and $\mathcal{R}_4$, have terms containing four structure constants, so all additional graphs have four nodes. There are two classes of graphs, depending on whether each node is connected to two others (``tin cans") or to three (``tetrahedra"). $\mathcal{R}_2$ contains only tin cans while $\mathcal{R}_4$ has both types of graphs. 

Tin can terms from the product of four Christoffel symbols have a symmetry structure that helps to simplify the large number of terms. Consider, for example, the term 
\begin{equation}
    g^{ef}g_{\ell k}\Gamma_{fdb}\Gamma_{ace}\Gamma^{kdb}\Gamma^{ac\ell}=\frac{1}{4}g^{ef}g_{\ell k}(C_{fdb}-C_{dbf}-C_{bdf})(C^{kdb}-C^{dbk}-C^{bdk})\Gamma_{ace}\Gamma^{ac\ell}\;.
\end{equation}
Here there are two pairs of terms that cancel, among the nine terms coming from expanding $\Gamma_{fdb}\Gamma^{kdb}$, due to the anti-symmetry of the full contraction under permutations of indices. On the other hand, there are two pairs of terms that add, containing $C_{dbf}C^{dbk}$ and $C_{bdf}C^{bdk}$, for example. There are therefore three sets of nine diagrams that remain to be evaluated. For example, one such term is 
\begin{align}
   g^{ef}g_{\ell k}C_{dbf}C^{dbk}&\left[C_{ace}C^{ac\ell}-C_{acd}C^{c\ell a}-C_{ace}C^{\ell ca}\right.\\\nonumber
    &-C_{cea}C^{ac\ell}+C_{cea}C^{c\ell a}+C_{cea}C^{\ell ca}\\\nonumber
    &\left.-C_{eca}C^{ac\ell}+C_{eca}C^{c\ell a}+C_{cea}C^{\ell ca}\right]\subset g^{ef}g_{\ell k}\Gamma_{fdb}\Gamma_{ace}\Gamma^{kdb}\Gamma^{ac\ell}\,.
\end{align}
The nine terms in this sum can be represented diagramatically as:

\begin{center}
\begin{tikzpicture}
[roundnode/.style={circle, draw=blue!20, very thick, minimum size=5mm}]
    \node[roundnode,fill=pink!20] (x1) {};
    \node[roundnode,fill=pink!20] (x2) [right=1.2cm of x1] {};
    \node[roundnode,fill=pink!20] (x3) [below=1.2cm of x2] {};
    \node[roundnode,fill=pink!20] (x4) [below=1.2cm of x1] {};

    \draw[dashed, ultra thick] (x1) -- (x2) {};
    \draw[-, ultra thick] (x1.250) -- (x4.110) {};
    \draw[dashed, ultra thick] (x1.290) -- (x4.70) {};
    \draw[-, ultra thick] (x2.250) --  (x3.110);
     \draw[dashed, ultra thick] (x2.290) -- (x3.70) {};
    \draw[dashed, ultra thick] (x3.west) -- (x4.east);

 \node[roundnode,fill=olive!20] (y1) [right=1.5cm of x2] {};
    \node[roundnode,fill=olive!20] (y2) [right=1.2cm of y1] {};
    \node[roundnode,fill=olive!20] (y3) [below=1.2cm of y2] {};
    \node[roundnode,fill=olive!20] (y4) [below=1.2cm of y1] {};

    \draw[dashed, ultra thick] (y1) -- (y2) {};
    \draw[-, ultra thick] (y1.250) -- (y4.110) {};
    \draw[dashed, ultra thick] (y1.290) -- (y4.70) {};
    \draw[-, ultra thick] (y2.250) edge[put token, postaction={decorate}] (y3.110);
     \draw[-, ultra thick] (y3.70) edge[put token, postaction={decorate}] (y2.290) {};
    \draw[dashed, ultra thick] (y3.west) -- (y4.east);

\node[roundnode,fill=green!20] (z1) [right=1.5cm of y2] {};
    \node[roundnode,fill=green!20] (z2) [right=1.2cm of z1] {};
    \node[roundnode,fill=green!20] (z3) [below=1.2cm of z2] {};
    \node[roundnode,fill=green!20] (z4) [below=1.2cm of z1] {};

    \draw[-, ultra thick] (z2) edge[put token, postaction={decorate}] (z1) {};
    \draw[-, ultra thick] (z1.250) -- (z4.110) {};
    \draw[dashed, ultra thick] (z1.290) -- (z4.70) {};
     \draw[-, ultra thick] (z3.70) edge[put token, postaction={decorate}] (z2.290) {};
     \draw[dashed, ultra thick] (z2.250) -- (z3.110);
    \draw[dashed, ultra thick] (z3.west) --  (z4.east);
    
\end{tikzpicture}

\vspace{0.5cm}

\begin{tikzpicture}
[roundnode/.style={circle, draw=blue!20, very thick, minimum size=5mm}]
    \node[roundnode,fill=olive!20] (x1) {};
    \node[roundnode,fill=olive!20] (x2) [right=1.2cm of x1] {};
    \node[roundnode,fill=olive!20] (x3) [below=1.2cm of x2] {};
    \node[roundnode,fill=olive!20] (x4) [below=1.2cm of x1] {};

    \draw[dashed, ultra thick] (x1) -- (x2) {};
    \draw[-, ultra thick] (x1.250) -- (x4.110) {};
    \draw[dashed, ultra thick] (x1.290) -- (x4.70) {};
    \draw[-, ultra thick] (x2.250) edge[put token, postaction={decorate}] (x3.110);
     \draw[-, ultra thick] (x3.70) edge[put token, postaction={decorate}] (x2.290) {};
    \draw[dashed, ultra thick] (x3.west) -- (x4.east);

 \node[roundnode,fill=pink!20] (y1) [right=1.5cm of x2] {};
    \node[roundnode,fill=pink!20] (y2) [right=1.2cm of y1] {};
    \node[roundnode,fill=pink!20] (y3) [below=1.2cm of y2] {};
    \node[roundnode,fill=pink!20] (y4) [below=1.2cm of y1] {};

    \draw[dashed, ultra thick] (y1) -- (y2) {};
    \draw[-, ultra thick] (y1.250) -- (y4.110) {};
    \draw[dashed, ultra thick] (y1.290) -- (y4.70) {};
    \draw[-, ultra thick] (y2.250) --  (y3.110);
     \draw[dashed, ultra thick] (y2.290) -- (y3.70) {};
    \draw[dashed, ultra thick] (y3.west) -- (y4.east);

    \node[roundnode,fill=green!20] (z1) [right=1.5cm of y2] {};
    \node[roundnode,fill=green!20] (z2) [right=1.2cm of z1] {};
    \node[roundnode,fill=green!20] (z3) [below=1.2cm of z2] {};
    \node[roundnode,fill=green!20] (z4) [below=1.2cm of z1] {};

     \draw[-, ultra thick] (z2) edge[put token, postaction={decorate}] (z1) {};
    \draw[-, ultra thick] (z1.250) -- (z4.110) {};
    \draw[dashed, ultra thick] (z1.290) -- (z4.70) {};
     \draw[-, ultra thick] (z3.70) edge[put token, postaction={decorate}] (z2.290) {};
     \draw[dashed, ultra thick] (z2.250) -- (z3.110);
    \draw[dashed, ultra thick] (z3.west) --  (z4.east);
    
\end{tikzpicture}

\vspace{0.5cm}

\begin{tikzpicture}
[roundnode/.style={circle, draw=blue!20, very thick, minimum size=5mm}]
    \node[roundnode,fill=green!20] (x1) {};
    \node[roundnode,fill=green!20] (x2) [right=1.2cm of x1] {};
    \node[roundnode,fill=green!20] (x3) [below=1.2cm of x2] {};
    \node[roundnode,fill=green!20] (x4) [below=1.2cm of x1] {};

    \draw[dashed, ultra thick] (x1) -- (x2) {};
    \draw[-, ultra thick] (x1.250) -- (x4.110) {};
    \draw[dashed, ultra thick] (x1.290) -- (x4.70) {};
    \draw[-, ultra thick] (x2.290) edge[put token, postaction={decorate}] (x3.70);
     \draw[dashed, ultra thick] (x3.110) -- (x2.250) {};
    \draw[-, ultra thick] (x3.west) edge[put token, postaction={decorate}] (x4.east);

 \node[roundnode,fill=green!20] (y1) [right=1.5cm of x2] {};
    \node[roundnode,fill=green!20] (y2) [right=1.2cm of y1] {};
    \node[roundnode,fill=green!20] (y3) [below=1.2cm of y2] {};
    \node[roundnode,fill=green!20] (y4) [below=1.2cm of y1] {};

    \draw[dashed, ultra thick] (y1) -- (y2) {};
    \draw[-, ultra thick] (y1.250) -- (y4.110) {};
    \draw[dashed, ultra thick] (y1.290) -- (y4.70) {};
    \draw[-, ultra thick] (y2.290) edge[put token, postaction={decorate}] (y3.70);
     \draw[dashed, ultra thick] (y3.110) -- (y2.250) {};
    \draw[-, ultra thick] (y3.west) edge[put token, postaction={decorate}] (y4.east);

    \node[roundnode,fill=red!40] (z1) [right=1.5cm of y2] {};
    \node[roundnode,fill=red!40] (z2) [right=1.2cm of z1] {};
    \node[roundnode,fill=red!40] (z3) [below=1.2cm of z2] {};
    \node[roundnode,fill=red!40] (z4) [below=1.2cm of z1] {};

     \draw[-, ultra thick] (z2) edge[put token, postaction={decorate}] (z1) {};
    \draw[-, ultra thick] (z1.250) -- (z4.110) {};
    \draw[dashed, ultra thick] (z1.290) -- (z4.70) {};
     \draw[dashed, ultra thick] (z3.70) -- (z2.290) {};
     \draw[dashed, ultra thick] (z2.250) -- (z3.110);
    \draw[-, ultra thick] (z3.west) edge[put token, postaction={decorate}] (z4.east);
    
\end{tikzpicture}
\end{center}

As the graph structure (and colors) indicate, the nine terms contain only four distinct constructions. Note that the graph structure does not directly indicate the relative sign between graphs related by a rotation of the structure nodes, for example between the upper right diagram and the lower left diagram. Furthermore, the signs of the terms are inherited from the signs in the expansions $\Gamma_{abc}=\frac{1}{2}(C_{abc}-C_{bca}-C_{cba})$. Collecting all relative signs, the two identical (green) diagrams in the right column cancel, as do the two identical (green) diagrams in the bottom row. The other pairs add so that there are three distinct diagrams. Within the four-gamma term that contains the above set, $g^{ef}g_{\ell k}\Gamma_{fdb}\Gamma_{ace}\Gamma^{kdb}\Gamma^{ac\ell}$, there are two other sets of nine terms that must be evaluated. The pattern of colors (identical graphs) and cancellations repeats in each of the nine sets of nine graphs, and there are ultimately three additional types of diagrams.

A different contraction structure occurs in terms like $g^{ef}g_{\ell k}C_{fcd}\Gamma_{aeb}\Gamma^{kdb}\Gamma^{ac\ell}$. This term can be represented by tetrahedron diagrams. Among the (seven) terms that do not cancel, there are just three diagrams:

\begin{center}
\begin{tikzpicture}
[roundnode/.style={circle, draw=blue!20, very thick, minimum size=5mm}]
    \node[roundnode,fill=violet!40] (x1) {};
    \node[roundnode,fill=violet!40] (x2) [right=1.2cm of x1] {};
    \node[roundnode,fill=violet!40] (x3) [below=1.2cm of x2] {};
    \node[roundnode,fill=violet!40] (x4) [below=1.2cm of x1] {};

    \draw[-, ultra thick] (x1) -- (x2) {};
    \draw[dashed, ultra thick] (x1.270) -- (x4.90) {};
    \draw[dashed, ultra thick] (x1.315) -- (x3.135) {};
    \draw[dashed, ultra thick] (x2.215) -- (x4.45);
     \draw[dashed, ultra thick] (x3.90) -- (x2.270) {};
    \draw[-, ultra thick] (x3.west) -- (x4.east);

 \node[roundnode,fill=red!40] (y1) [right=1.5cm of x2] {};
    \node[roundnode,fill=red!40] (y2) [right=1.2cm of y1] {};
    \node[roundnode,fill=red!40] (y3) [below=1.2cm of y2] {};
    \node[roundnode,fill=red!40] (y4) [below=1.2cm of y1] {};

    \draw[-, ultra thick] (y1) edge[put token, postaction={decorate}] (y2) {};
    \draw[dashed, ultra thick] (y1.270) -- (y4.90) {};
   \draw[dashed, ultra thick] (y1.315) -- (y3.135) {};
    \draw[dashed, ultra thick] (y2.215) -- (y4.45);
     \draw[-, ultra thick] (y3.90) -- (y2.270) {};
    \draw[-, ultra thick] (y4.east) edge[put token, postaction={decorate}] (y3.west);

    \node[roundnode,fill=blue!40] (z1) [right=1.5cm of y2] {};
    \node[roundnode,fill=blue!40] (z2) [right=1.2cm of z1] {};
    \node[roundnode,fill=blue!40] (z3) [below=1.2cm of z2] {};
    \node[roundnode,fill=blue!40] (z4) [below=1.2cm of z1] {};

     \draw[-, ultra thick] (z2) edge[put token, postaction={decorate}] (z1) {};
    \draw[-, ultra thick] (z1.270) edge[put token, postaction={decorate}] (z4.90) {};
   \draw[dashed, ultra thick] (z1.315) -- (z3.135) {};
    \draw[dashed, ultra thick] (z2.215) -- (z4.45) {};
     \draw[-, ultra thick] (z3.90) edge[put token, postaction={decorate}] (z2.270) {};
    \draw[-, ultra thick] (z4.east) edge[put token, postaction={decorate}] (z3.west);
    
\end{tikzpicture}
\end{center}

One can certainly consider other loss functionals. For example, the function primarily studied in \cite{Freedman:2021a}, given by Eq.(\ref{eq:FZNGloss}), contains terms with all even numbers of $C_{ijk}$. That work considered an expansion of terms up to third order, giving rise to graphs with two, four, and six nodes. However, only a subset of the four-vertex graphs appeared. The set of four-vertex graphs needed for the curvature terms is larger.

\end{document}